\documentclass{article}  
\usepackage{amssymb}

\newcommand{\ul}[1]{\underline{#1}}

\begin{document}

\begin{center}
{\Large \bf Vertical transport and domain formation in
multiple quantum wells}\\[0.3cm]

{\Large Andreas Wacker}\\[0.1cm]
{\em
Mikroelektronik Centret,
Danmarks Tekniske Universitet, DK-2800 Lyngby, Denmark\\
email: {\tt wacker@mic.dtu.dk}}

\vspace{2cm}

to appear in:\\[0.1cm]
 {\bf Theory of Transport Properties of
Semiconductor Nanostructures}\\[0.1cm]
edited by Eckehard Sch{\"o}ll\\
 (Chapman and Hall, in preparation)
\end{center}
\vspace{1cm}
\begin{quotation}
In this book article effects related to the vertical transport in
weakly coupled multiple quantum wells are reviewed.
A self-contained microscopical  model for the calculation of the
well-to-well currents without any fittings parameters is presented.
The model exhibits the well-known peaks in the current-field relation
in quantitative agreement with experiments.
This local current-field relation is used as an input for the calculation
of the transport in the extended structure consisting of  many periods.
Here both the formation of stationary field domains as well as
self-sustained current oscillations are found in good agreement with
experimental data. The underlying physics of these nonlinear phenomena
is discussed in detail.
\end{quotation}

\newpage
\setcounter{section}{-1}
\section{Introduction}
\begin{figure}[b]
\vspace*{3.4cm}
\includegraphics{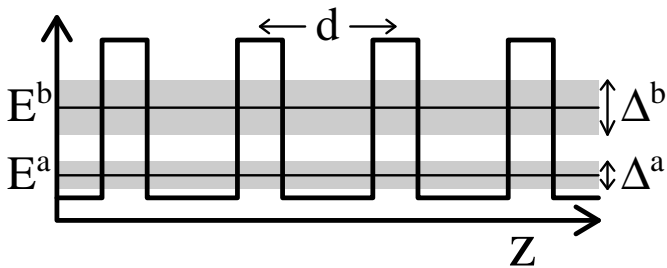}
\caption[a]{Sketch of the conduction band $E_c(z)$ with minibands $\nu=a,b$} 
\label{Figminiband}
\end{figure}

Today's growth techniques allow the construction of
semiconductor structures where layers of different semiconductors
(exhibiting similar lattice constants) can be grown on each other 
with the interface being well defined within one atomic monolayer.
If two such layers alternate periodically one obtains a periodic structure
with an artificial  period $d$ in the growth direction (which is defined 
to be the $z$-direction). This leads to spatial variations 
in the conduction and valence
band of the material as sketched in Fig.~\ref{Figminiband}.
Considering only conduction band states in the following, 
the region with lower conduction band is called 'well' and the region with
higher conduction band 'barrier'.
An extended discussion of various aspects of such structures 
can be found in Ref.~\cite{GRA95d}.
In this chapter the vertical electronic transport (i.e., in $z$-direction)
is considered for such structures.

If we consider the electronic properties  of such a structure,
there are two different approaches.
In the first approach one may consider the full structure as an artificial
lattice. The energy spectrum can be calculated
analogously to the Kronig-Penney model (which is discussed in
almost all solid-state physics textbooks)  resulting in the appearance of
energy bands and energy gaps as sketched in Fig.~\ref{Figminiband}. 
Due to this analogy with the atomic lattice of lattice constant $a_L$ 
such semiconductor structures are often called superlattices.
The corresponding eigenfunctions are the usual Bloch functions 
which extend over the whole superlattice.
The Bloch functions are labeled with the Bloch vector $q$ which is
restricted to the Brillouin zone $-\pi/d<q<\pi/d$. This range 
is  much smaller than the  Brillouin zone $-\pi/a_L<q<\pi/a_L$ of the atomic 
lattice as $d\gg a_L$. Therefore these new bands are
called minibands. If an electric field is applied,
the Bloch functions are accelerated and the transport
can be treated analogously to the usual transport in bulk systems \cite{ESA70}.
Due to the short Brillouin zone many special effects like Bloch oscillations 
can be found here which are not accessible in bulk systems as
discussed in the preceding chapter.  

In the second approach one may consider the wells to be isolated 
from each other as a first approximation and calculate the 
eigenstates within each well. This yields a sequence of energy 
levels and localized wave functions for each well.
Such a structure is usually called a multiple quantum well.
Of course this approach only makes sense if the coupling between different wells
is weak and can be calculated perturbatively. 
Thinking in terms of Fermi's golden rule, there will be 
transitions between the levels in different wells
yielding sequential tunnelling. 
Due to energy conservation this tunnelling takes only place if
the energy levels align and strong resonances \cite{CAP86} are likely to occur
if the relative height of the levels in different wells
is varied by a voltage applied in the $z$-direction.

The width of these resonances is related to scattering processes of the 
electrons inside the wells. In Ref.~\cite{MUR95} this feature was used to
determine the electron-electron scattering rate from the
transport between two weakly coupled quantum wells. This shows
that a  modelling of scattering processes is necessary
for a calculation of the current in such structures and
opens the possibility for checking theoretical concepts as well as
investigating the significance of various scattering mechanisms
by comparison with experimental data.

These resonances yield strong nonlinearities in the local
current-field relation. If we now have a long periodic sequence
of these coupled quantum wells the full system is a good example of an 
extended nonlinear system. This leads to the appearance
of complicated current-voltage characteristics exhibiting an almost periodic
sequence of branches
due to the formation of electric field domains as well as to self-sustained
oscillations due to moving domains.
As the periodicity of the structure is an important feature for the 
translational invariance, these structures are frequently  called
superlattices for weak  coupling as well.
If the translational invariance is broken due to an insufficient control
of the growth conditions the branches lose their periodic structure,
and  information about the actual structure can be obtained 
from measurements of the current-voltage characteristics.

The first section of this chapter is devoted to the classification
of the different regimes depending on the coupling between the wells.
In the second section  a  transport  model for weakly
coupled quantum wells is presented
which allows a microscopic calculation of the current without
any fitting parameters. In the third chapter this model is extended in order
to describe domain formation. The fourth section is on
the influence of deviations from the periodic structure of 
the superlattice and the fifth section refers to self-sustained
oscillations.
Finally, details of the calculations will be given in the sixth section.

\section{The different transport regimes}
In this section  the precise meaning of weakly and strongly
coupled quantum wells is discussed.

Assuming ideal interfaces the semiconductor 
structure is translational invariant within
the $x$ and $y$ direction perpendicular to the growth direction.
Therefore the $x,y$ dependence can be separated by
the ansatz $e^{i\ul{k}\cdot \ul{r}}$ where
$\ul{k}$ and $\ul{r}$ are vectors within the two-dimensional
$(x,y)$ plane. 
(The effects  due to the atomic periodicity of the
lattice are treated  within the envelope function formalism \cite{BAS88}
assuming a parabolic subband.)
Within the $z$-direction we have an artificial period $d$
leading to energy states $E^{\nu}_q$ 
characterized by the miniband index  $\nu$ and the 
quasi-wave vectors $-\pi/d<q\le \pi/d$. 
The energies as well as the wave-functions $\varphi^{\nu}_q(z)$
can be calculated numerically analogously to the well-known
Kronig-Penney model. 
For a given miniband $\nu$ the energy $E^{\nu}_q$ has a mean value
$E^{\nu}=d/(2\pi)\int_{-\pi/d}^{\pi/d}{\rm d}q E^{\nu}_q$ and varies
within the miniband width 
$\Delta^{\nu}={\rm Max}_q(E^{\nu}_q)-{\rm Min}_q(E^{\nu}_q) $ 
as sketched in Fig.~\ref{Figminiband}.

\begin{figure}
\vspace*{5.1cm}
\includegraphics{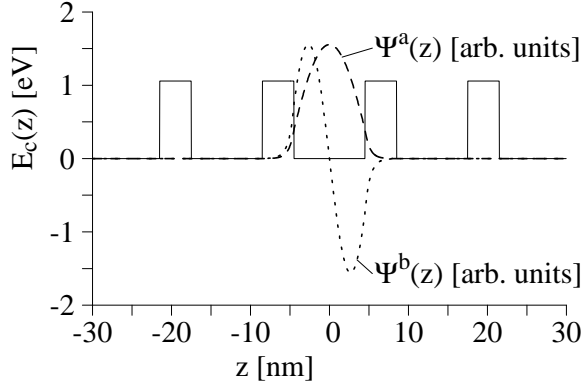}
\caption[a]{Conduction band $E_c(z)$ together
with the Wannier functions calculated for the two lowest subbands.}
\label{Figwannier}
\end{figure}

From the Bloch functions $\varphi_q^{\nu}(z)$ the Wannier
functions $\Psi^{\nu}(z-nd)$ defined by
\begin{eqnarray}
\Psi^{\nu}(z-nd)=\sqrt{\frac{d}{2\pi}}\int_{-\pi/d}^{\pi/d} {\rm d}q 
e^{-inqd} \varphi_q^{\nu}(z)\label{Eqwannier}
\end{eqnarray} 
can be constructed.
They are real and localized in well $n$ for an appropriate choice
of the complex phase for $\varphi_q^{\nu}(z)$ \cite{KOH59}. 
An example is shown in Fig.~\ref{Figwannier}.
Using the Fourier expansion
\begin{equation} 
E^{\nu}_q=E^{\nu}
+\sum_{h=1}^{\infty} 2 T^{\nu}_h  \cos(hdq)\label{Eqfourier}
\end{equation} 
one obtains the following Hamiltonian in second quantization
\begin{equation}
\hat{H}=\sum_{n,\nu} 
E^{\nu}c^{\nu\dag}_nc^{\nu}_n
+\sum_{h=1}^{\infty}\left( T^{\nu}_h 
c^{\nu\dag}_{n+h}c^{\nu}_n+T^{\nu}_hc^{\nu\dag}_{n-h}c^{\nu}_n \right)
\label{Eqhamintro}
\end{equation} 
for the creation  $c^{\nu\dag}_n$ and annihilation $c^{\nu}_n$ operators 
of the Wannier functions $\Psi^{\nu}(z-nd)$.
As the Wannier functions are linear combinations of Bloch functions
with different energies, they are no stationary states.
Neglecting terms with $h>1$ the time evolution of the annihilation operators
is given by
\begin{equation}
i\hbar \frac{{\rm d}}{{\rm d}t}c^{\nu}_n=
T_1^{\nu}\left(c^{\nu}_{n+1}+ c^{\nu}_{n-1}\right)\label{Eqcoherent}\, .
\end{equation}
For the initial condition $c^{\nu}_n(t=0)=\delta_{n,0}c^{\nu}_0$ this
has the solution 
\begin{equation}
c^{\nu}_n(t)=i^{-n}J_n\left(\frac{2T^{\nu}_1}{\hbar}t\right)c^{\nu}_0
\end{equation} 
where $J_n(x)$ is the Bessel function of order $n$.
Since $\sum_{n=-\infty}^{\infty}n^2(J_n(x))^2=x^2/2$
the average extension $n_a(t)$ as a function of time is found to be
\begin{equation}
n_a(t)=\sqrt{\langle n^2 \rangle}=\frac{\sqrt{2}T_1}{\hbar}t\, .
\end{equation}
This is obviously a coherent process. In a real semiconductor structure
there will be scattering (characterized by a scattering time $\tau_{\rm sc}$)
destroying  the coherent evolution\footnote{This scattering can be
a phase breaking process such as scattering with phonons.
But also elastic scattering processes at impurities or interface
roughness will contribute here because they destroy the
translational invariance of the superlattice, which is both
the justification for the existence of the Bloch functions and a WS state.}
given by Eq.~(\ref{Eqcoherent}). 
If $n_a(\tau_{\rm sc})\ll 1$ holds,
the  phase coherence is completely lost during the spread and
the states of adjacent wells  will not maintain a
fixed phase relation. In this case the transitions can be
described by sequential tunnelling between the Wannier states 
with loss of phase between different tunnelling events.
Defining $\Gamma=\hbar/\tau_{\rm sc}$ one therefore may use the
picture of 
\begin{equation}
\fbox{sequential tunnelling for $T_1\ll \Gamma/\sqrt{2}$ .} 
\end{equation}

If, on the other hand, the extended Bloch functions shall be a reasonable 
basis set, the phase coherence has to be maintained over a 
larger number of quantum wells.
This gives $n_a(\tau_{\rm sc})\gg 1$ as a first necessary condition.
But a further complication arises if an electric field $F$ is applied
in the $z$ direction. As discussed in the preceding chapter of this book
the eigenvalues of the Hamiltonian take discrete values  $E^{\nu}-neFd$ 
which is called the Wannier-Stark (WS) ladder.
The corresponding eigenfunctions $\phi^{\nu}_{\rm  WS}(z-nd)$ are
localized around well $n$. Within the one band limit
they are given by
\begin{equation}
\phi^{\nu}_{\rm WS}(z)=\sum_n J_{n}\left(\frac{2T^{\nu}_1}{eFd}\right)
\Psi^{\nu}(z-nd)\, .
\end{equation}
(See, e.g., Ref.~\cite{EMI87}, where also the coupling between different
bands is discussed.)
Then the spatial extension of the WS-states is given by 
\begin{equation}
n_{\rm WS}=\sqrt{\langle n^2 \rangle}=\frac{\sqrt{2}T_1}{eFd}\, .
\label{EqWSextension}
\end{equation}
If now $n_{\rm WS}$ becomes small, the electrons are localized and the
use of extended Bloch functions does not make any sense.
This provides the second condition $n_{\rm WS}\gg 1$ for
the Bloch functions to be an appropriate basis set of wave functions.
Thus it  only makes sense to speak about
\begin{equation}
\fbox{
minibands for $T_1\gg \Gamma/\sqrt{2}$ and
$T_1\gg eFd\sqrt{2}$ .\label{Eqcondminiband}}
\end{equation}

As a third basis set the Wannier-Stark functions $\phi^{\nu}_{\rm WS}(z-nd)$ 
may be used if an electric field is applied. They make sense if coherence is
maintained within their extension, i.e., if $n_{\rm WS}<n_a(\tau_{\rm sc})$.
Therefore the transport can be described by
\begin{equation}
\fbox{WS hopping for $eFd>\Gamma$ .}
\end{equation}

\begin{figure}
\vspace*{5.6cm}
\includegraphics{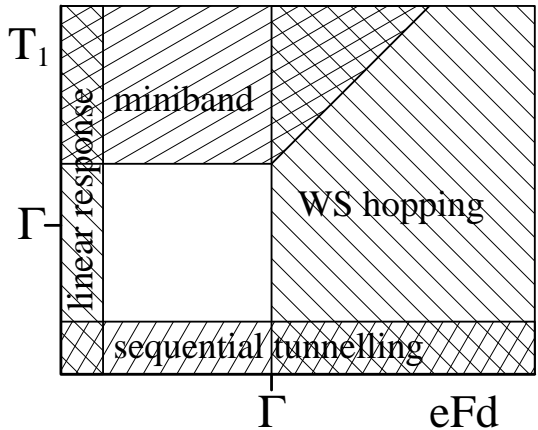}
\caption[a]{Sketch of the regimes where different approaches are valid} 
\label{Figregimes}
\end{figure}

These different regimes of validity are sketched in Fig.~\ref{Figregimes}
where the condition $a\ll b$ has been translated to $a<b/2$ for illustrative
purpose.

How can the electric transport be described if
an electric field $F$  is applied in the $z$ direction?
For $F=0$ the system should be in equilibrium and no current $I$ flows.
For small fields the current should increase with $F$. 
(This may be calculated by
linear response which is given by the Kubo formula for a quantum system.
See, e.g., chapter 5, where the application of the Kubo formula 
has been described  for the Coulomb drag.)
Now it is an important feature of both superlattices
and multiple quantum wells that for larger fields
there is a range with negative differential conductivity 
${\rm d}I/{\rm d}F<0$ (NDC). This can be easily understood
within the Wannier-Stark picture.
As the WS-states are orthogonal, transitions between them have to be
caused by scattering events. Now the matrix element
for a given scattering element depends on the spatial overlap
of the eigenfunctions.  With increasing field
the Wannier-Stark states become more and more localized
as can be seen from Eq.~(\ref{EqWSextension}). Therefore
the matrix element for the scattering is decreased and the
hopping rate is diminished \cite{TSU75,ROTp}.
As the WS approach is not justified for low electric fields
(see Fig.~\ref{Figregimes}) an Ohmic behaviour for low fields
could not be observed within this model \cite{ROTp}.

In the  validity range of 
(\ref{Eqcondminiband}) the basis of Bloch functions may be used.
Due to an electric field the Bloch waves are accelerated according to
$\hbar \dot{q}=eF$. Like in the standard transport theory for
bulk systems  a positive conductivity 
is found for low electric fields.
But for higher fields the Bloch vector $q$ may run through the whole
Brillouin zone $2\pi/d$ if $eF\tau_{\rm sc}/\hbar >2\pi/d$ holds.
Then the electrons perform a periodic motion in both $q$ and $z$ space
which is called the Bloch oscillation. A quantitative 
analysis within various approaches \cite{ESA70,LEB70,IGN91,LEI91}
reveals 
\begin{equation}
I\sim \frac{eFd}{(eFd)^2+\Gamma^2}\label{Eqcurrentmini}\, .
\end{equation}
This yields positive differential conductivity for $eFd<\Gamma$
and negative differential conductivity for larger fields which
is confirmed by experimental data (see, e.g., \cite{SIB90,SCH96e}).
For large $F$ there is $I\sim 1/F$. This behaviour is also found
from the WS-hopping model for $eFd<\Delta=4T_1$ \cite{ROTp} which
can be easily understood from Fig.~\ref{Figregimes}
as this condition includes the  region where both approaches are valid.
This is an explicit example of the more general equivalence
between the Wannier-Stark picture and the Bloch oscillation picture
discussed in chapter 9 of this book.

In case of weakly coupled multiple quantum wells  $T_1\ll \Gamma/\sqrt{2}$ 
NDC has been observed experimentally as well \cite{GRA91a}.
Here we can consider the electronic states to be the levels of the single
quantum wells which will exhibit a certain width $\Gamma$ due to scattering. 
The tunnelling between the wells is caused by the coupling
$T_1$. Due to energy conservation a significant current between 
adjacent wells may flow if the levels are 
aligned within accuracy $\Gamma$. Therefore NDC is likely to
appear if the disalignment $eFd$ becomes larger than $\Gamma$ 
\cite{KAZ72,MIL94}. 
This will be modelled quantitatively in the next section.

For the white region of Fig.~\ref{Figregimes} between these two limits
the situation is more complicated.
Here Laikhtman and Miller \cite{LAI93} were able 
to show that Eq.~(\ref{Eqcurrentmini})
holds provided the electron temperature is much larger than
$T_1$ and $eFd$.

In an extended system the existence of an NDC region
typically causes oscillatory behaviour due to travelling
field domains (see, e.g., \cite{SHA92}). For the miniband regime
this has been  studied theoretically quite early in 
Refs.~\cite{BUE77,BUE79,IGN85} and recently also 
experimentally \cite{PER92,HOF96}.
Note that these travelling domain oscillations 
are self-sustained in Ref.~\cite{HOF96} while for the usual
Bloch oscillations only transient oscillations are found 
(see, e.g., Ref.~\cite{WAS93}).
The same type of oscillations has been recently found
in the regime of sequential tunnelling both experimentally 
\cite{KAS95,GRA96,OHT96} and by numerical simulations \cite{BON95,WAC95}.
An analytic treatment of the oscillation mode is given
in Refs.~\cite{KAS97,BON97}.
Under certain conditions chaotic oscillations occur as well \cite{BUL95,ZHA96}.

For weakly coupled multiple quantum wells an additional  scenario 
occurs, which has been extensively studied in the last decade: 
As the electronic states are localized within single wells the
domain boundary (which are charge accumulation layers) can be
trapped in a single quantum well and  a stationary stable domain is formed.
As the domain boundary may be located in any well a periodic sequence 
of branches appears in the current-voltage characteristic.
This effect has been observed experimentally by many groups
\nocite{ESA74,KAW86,CHO87,HEL89,HEL90,GRA91,MUR94,KAS94,ZHA94,KWO95,STO95,
HAN95a,KEA95a,ZEU96,MIM96}\cite{ESA74}- \cite{MIM96}.
Theoretically such effects can be studied by combining rate equations between
the wells and Poisson's equation.
Such an approach has already been performed in
Refs.~\cite{SUR74,LAI91}. Nevertheless, to my knowledge,
the full current-voltage characteristic
exhibiting the domain structure could first  be resolved qualitatively in
Ref.~\cite{KOR93} for slim superlattices exhibiting a few electrons per well
and in Refs.~\cite{PRE94,BON94} for two-dimensional wells. 
A full quantitative calculation has been
presented recently \cite{WACpa,WACpb} for different multiple quantum wells
yielding quantitative agreement with experimental data without
using any fitting parameters.
This model will be presented in the next sections.

\section{Modelling of the transport between weakly coupled  quantum wells}
In this section I want to show how the currents
between the wells can be calculated from a microscopic model.
Furthermore I present some  simplifications in order to obtain 
estimates for the current.
The numerical calculations are performed  with the data of the
sample used in Refs.~\cite{KAS95,GRA91} exhibiting $N=40$ GaAs wells of width 
$w=9$ nm between 41 AlAs barriers of width $b=4$ nm.
The wells are n-doped with a doping density of 
$N_D=1.5\times 10^{11}/{\rm cm}^2$ per well.
Details of the specific calculations outlined in this
section are given in section \ref{Secdetails}.

As discussed in the first section, for weakly coupled quantum wells
the products of Wannier functions $\Psi^{\nu}(z-nd)$, 
localized in well $n$, and plane waves $e^{i\ul{k}\cdot \ul{r}}$
form a  reasonable basis set of wave functions. 
Now we restrict ourselves to  the lowest two levels denoted by $\nu=a,b$.
The respective annihilation operators are denoted by
$a_n=c^{a}_{n}$ and $b_n=c^{b}_{n}$.
Furthermore we add the contribution $-eFz$ due to a homogeneous 
electric field $F$ in the Hamiltonian (\ref{Eqhamintro})
where $e<0$ is the charge of the electron.
Restricting ourselves to coupling between neighbouring wells ($h=1$) 
we obtain the Hamiltonian $\hat{H}=\hat{H}_0+\hat{H}_1+\hat{H}_2$:
\begin{eqnarray}
\hat{H}_0^{{\rm res}}&=&\sum_{n,\ul{k}} \left[
(E^{a}+E_k-eFdn)a_n^{\dag}(\ul{k})a_n(\ul{k})\right.\nonumber \\
&&\left. +(E^{b}+E_k-eFdn)b_n^{\dag}(\ul{k})b_n(\ul{k})
\right]\label{Eqham1}\\
\hat{H}_1^{{\rm res}}&=&\sum_{n,\ul{k}} \left[
T_1^a a_{n+1}^{\dag}(\ul{k})a_n(\ul{k})
+T_1^b b_{n+1}^{\dag}(\ul{k})b_n(\ul{k})\right.\nonumber \\
&&\left. -eFR^{ab}_1 a_{n+1}^{\dag}(\ul{k})b_n(\ul{k})
-eFR^{ba}_1 b_{n+1}^{\dag}(\ul{k})a_n(\ul{k}) 
+h.c.\right]\label{Eqham2}\\
\hat{H}_2^{{\rm res}}&=&\sum_{n,\ul{k}}
\left[ -eF(R_0^{ab}a_n^{\dag}(\ul{k})b_n(\ul{k})
+R_0^{ba}b_n^{\dag}(\ul{k})a_n(\ul{k}))\right]
\label{Eqham3}
\end{eqnarray}
with the parabolic dispersion $E_k=\hbar^2k^2/(2m_w)$ 
($m_w$ is  the effective mass  in the well) and  the couplings 
$R_h^{\nu'\nu}=\int dz \Psi^{\nu'}(z-hd)z\Psi^{\nu}(z)$.
All energies $E$ are given with respect to the bottom
of the quantum well.
The values of the coefficients are presented in Table~\ref{Tabcoeff}.
The term $\hat{H}_2$ can be incorporated into the one electron states 
by diagonalizing $\hat{H}_0+\hat{H}_2$ \cite{KAZ72}. 
This leads to renormalized field-dependent
coefficients in $\hat{H}_0$ and $\hat{H}_1$ (which are used in the following) 
but does not change the structure of the problem for a homogeneous 
electric field.

\begin{table}
\begin{tabular}{|l|l|l|}
\hline
$E^a=47.1$ meV & $T_1^a=-0.0201$ meV & $R_0^{ba}=-0.149d$\\ \hline
$E^b=$ 176.6 meV & $T_1^b=0.0776$ meV & $R_1^{ba}=2.66\times 10^{-4}d$
\\ \hline
\end{tabular}
\caption[a]{Calculated level energies and transition elements
for Eqs.~(\ref{Eqham1}-\ref{Eqham3}).}
\label{Tabcoeff}
\end{table}

$\hat{H}^{{\rm res}}$ is only considering $\ul{k}$-conserving
processes reflecting an ideal structure with translational 
invariance in the $\ul{r}$-plane and neglecting
any many-particle processes.
Nevertheless, there are non $\ul{k}$-conserving processes as well
which may result from scattering at impurities or interface roughness,
e.g.. These give two  further contributions to the Hamiltonian: 
\begin{eqnarray}
&&\hat{H}^{{\rm scatter}}_0=\frac{1}{A}
\sum_{\ul{k},\ul{p}}\left[
U^{aa}_{0}(\ul{p})a^{\dag}_{n}(\ul{k}+\ul{p})a_{n}(\ul{k})
+U^{bb}_{0}(\ul{p})b^{\dag}_{n}(\ul{k}+\ul{p})b_{n}(\ul{k})\right.
\nonumber\\
&&\quad\left.+U^{ba}_{0}(\ul{p})b^{\dag}_{n}(\ul{k}+\ul{p})a_{n}(\ul{k})
+U^{ab}_{0}(\ul{p})a^{\dag}_{n}(\ul{k}+\ul{p})b_{n}(\ul{k})
\right] \label{EqH0scatter}
\end{eqnarray}
is the contribution due to scattering within the well and
\begin{eqnarray}
&&\hat{H}^{{\rm scatter}}_1=\frac{1}{A}
\sum_{\ul{k},\ul{p}}\left[
U^{aa}_{1}(\ul{p})a^{\dag}_{n+1}(\ul{k}+\ul{p})a_{n}(\ul{k})
+U^{bb}_{1}(\ul{p})b^{\dag}_{n+1}(\ul{k}+\ul{p})b_{n}(\ul{k})\right.
\nonumber\\
&&\quad \left.
+U^{ba}_{1}(\ul{p})b^{\dag}_{n+1}(\ul{k}+\ul{p})a_{n}(\ul{k})
+U^{ab}_{1}(\ul{p})a^{\dag}_{n+1}(\ul{k}+\ul{p})b_{n}(\ul{k})
+h.c.\right]
\end{eqnarray}
refers to interwell scattering, where the restriction to
neighboured wells is made.

Within the assumptions of local thermal equilibrium in each well
and weak coupling between the wells
the current from level $\nu$ in well $n$ to level 
$\mu$ in well $n+1$ is given
by the following expression\footnote{An important
point of the derivation is the assumption of  uncorrelated  scattering 
in different wells. For example, this is the fact if
the electrons are dominantly scattered by the impurities  localized
in the same well. If the scattering occurs at identical impurities
for the electrons in well $n$ and in well $n+1$ correlation effects
occur which may essentially change the result, see Refs.~\cite{ZHE93,KAZ72}.}  
(see section 9.3 of Ref.~\cite{MAH90}):
\begin{eqnarray}
\lefteqn{I_{n\to n+1}^{\nu \to \mu}=2e\sum_{\ul{k}',\ul{k}}
|H_{(n+1)\ul{k}',n\ul{k}}^{\mu,\nu}|^2
\int_{-\infty}^{\infty} \frac{dE}{2\pi \hbar}
A_n^{\nu}(\ul{k},E)}\cdot \label{EqJ} \\
&&\cdot A_{n+1}^{\mu}(\ul{k}',E+eFd)
\left[n_F(E-E_n^F)-n_F(E-E_{n+1}^F+eFd)\right]\, .
\nonumber 
\end{eqnarray}
Here $E_n^F$ is the chemical potential in well $n$ which is
measured with respect to the bottom of the quantum well.
$n_F(x)=(1+e^{x/(k_BT_e)})^{-1}$ is the Fermi function
and $T_e$ is the electron temperature.
$A_n^{\nu}(\ul{k},E)$ denotes the spectral function for the state
$\ul{k}$ of the subband $\nu$ in well number $n$.
It represents the weight of the free  particle
state  $\ul{k}$ contributing to the energy $E$. Then  the
total density of states $\rho_n^{\nu}(E)$ in subband $\nu$ is given by
\begin{equation}
\rho_n^{\nu}(E)=\frac{2}{2\pi A}\sum_{\ul{k}}A_n^{\nu}(\ul{k},E)
\label{Eqdichte}
\end{equation}
where the factor $2$ reflects the spin degeneracy and $A$ denotes the sample
area.
If no scattering is present, the state $\ul{k}$ has a fixed energy
$E^{\nu}+E_{k}$, and the spectral function becomes
a $\delta$-function $A_n^{\nu}(\ul{k},E)=2\pi 
\delta (E-E^{\nu}-E_{k})$. In the 
continuum limit ($\sum_{\ul{k}}\to A/(2\pi)^2\int d^2k$) 
Eq.~(\ref{Eqdichte}) gives then the
2-dimensional density of states $\rho_n^{\nu}(E)=\rho_0\theta(E-E^{\nu})$
with $\rho_0=\frac{m}{\pi\hbar^2}$.

If scattering is present due to $\hat{H}_0^{\rm scatter}$
the states $\ul{k}$ are no longer eigenstates
of the total Hamiltonian. This can be taken into account
by calculating the self-energy for the given scattering within
standard theory (see, e.g., \cite{MAH90}).
Assuming equilibrium  the spectral function is then related 
to the  retarded self-energy $\Sigma_{n}^{\nu\,{\rm ret}}(\ul{k},E)$ via 
\begin{equation}
A_n^{\nu}(\ul{k},E)=\frac{-2 {\rm Im}\Sigma_{n}^{\nu\,{\rm ret}}(\ul{k},E)}
{\left(E-E^\nu-E_k-
{\rm Re}\Sigma_{n}^{\nu\,{\rm ret}}(\ul{k},E)\right)^2
+\left({\rm Im}\Sigma_{n}^{\nu\,{\rm ret}}(\ul{k},E)\right)^2}\, .
\end{equation}
Frequently, the self-energy is taken to be constant for simplicity
setting
$\Sigma_{n}^{\nu\,{\rm ret}}(\ul{k},E)\approx W_{n}^{\nu}-i\Gamma_{n}^{\nu}/2$.
Then the spectral function is a Lorentzian with a full
width at half maximum $\Gamma$.

\begin{figure}
\vspace*{3.7cm}
\includegraphics{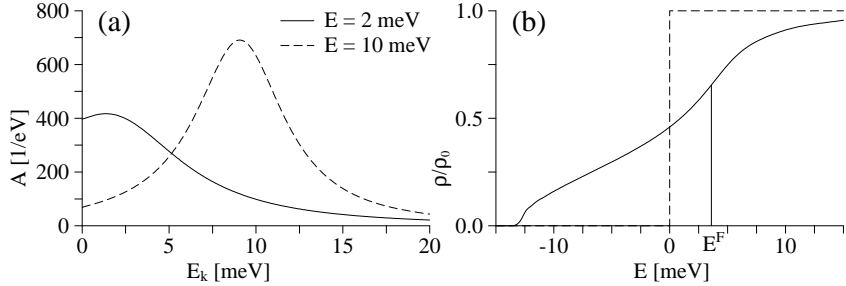}
\caption[a]{(a) Spectral function $A^a(E_k,E)$
of the lowest level for two different energies $E$ calculated for
impurity scattering and interface roughness (see section~\ref{Secdetails}).
(b) Density of states in units of the free-particle density of
states (dashed line) where we have also indicated the
value of the Fermi-energy at zero temperature.}
\label{Figspektral}
\end{figure}

Here the self-energies are calculated from basic scattering processes
at impurities and interface roughness without using any fitting parameters.
This calculation is presented in section~\ref{Secdetails}.
The calculated spectral function for different energies $E$ are shown in 
Fig.~\ref{Figspektral}(a) for illustration. One can clearly see
that they exhibit a maximum close to $E_k=E$. From the width we may estimate
that $\Gamma\approx10$ meV holds for $E=2$ meV and $\Gamma\approx$ 6 meV 
for $E=10$ meV. Fig.~\ref{Figspektral}(b) shows the total density of states.

While  the full derivation is slightly tedious \cite{MAH90}
formula (\ref{EqJ}) can be motivated quite easily:
In the long-time limit energy has to be conserved during transitions
caused by the time-independent interwell couplings
$H_{(n+1)\ul{k}',n\ul{k}}^{\mu,\nu}$.
Therefore we have to consider tunnelling processes for
a certain energy $E$
and integrate over $E$ afterwards.
The factor $[n_F(E-E_n^F)-n_F(E+eFd-E_{n+1}^F)]$ takes
into account the thermal occupation at the given energy in both wells.
The free particle state $\ul{k}$ has a weight
$A_n^{\nu}(\ul{k},E)/(2\pi)$ in well $n$.
Its transition probability to the state $\ul{k}'$ in well $n+1$ is
given by
$2\pi |H_{(n+1)\ul{k}',n\ul{k}}^{\mu,\nu}|^2/\hbar $ (Fermi's golden rule). 
The final state
has a weight $A_{n+1}^{\mu}(\ul{k}',E+eFd)/(2\pi)$
at the given energy.
Obviously one has to sum over all free particle states
$\ul{k},\ul{k}'$.
Finally, the factor 2 is due to the spin degeneracy.

\subsection{Resonant Transitions}
Let us first investigate the current due to the transition 
elements from $\hat{H}_1^{{\rm res}}$ in Eq.~(\ref{Eqham2}). 
They conserve the momentum $\ul{k}$ and therefore the kinetic energy $E_k$.
Using Eq.~(\ref{Eqdichte}) one can rewrite Eq.~(\ref{EqJ}) as
\begin{eqnarray}
I_{n\to n+1}^{\nu \to \mu,{\rm res}}&=&
A\frac{e|H_{1}^{{\rm res}\, \mu,\nu}|^2}{\hbar}
\int_{-\infty}^{\infty} dE \rho_n^{\nu}(E)\langle A_{n+1}^{\mu}\rangle(E,eFd)
\nonumber \\
&&\left[n_F(E-E_n^F)-n_F(E-E_{n+1}^F+eFd)\right]
\end{eqnarray}
with the average
\begin{equation}
\langle A_{n+1}^{\mu} \rangle (E,eFd)=
\frac{\sum_{\ul{k}}A_n^{\nu}(\ul{k},E) A_{n+1}^{\mu}(\ul{k},E+eFd)}
{\sum_{\ul{k}}A_n^{\nu}(\ul{k},E)}\, .
\end{equation}
If  the spectral functions are $\delta$-functions 
$\langle A_{n+1}^{\mu} \rangle (E,eFd)=2\pi \delta(E_{\nu}+eFd-E_{\mu})$ holds.
Therefore tunnelling only takes place if the levels are exactly aligned.
In order to estimate the effect of broadening one may assume 
constant self-energies $\Sigma_{n}^{\nu\,{\rm ret}}=-i\Gamma^{\nu}/2$.
Performing the continuum limit and assuming 
$E-E^{\nu}\gg \Gamma_{n}^{\nu}$ as well as 
$E+eFd-E^{\mu}\gg \Gamma^{\mu}$ one finds
\begin{equation}
\langle A_{n+1}^{\mu} \rangle (E,eFd)=
\frac{\Gamma^{\nu}+\Gamma^{\mu}}
{(eFd+E^{\nu}-E^{\mu})^2+(\Gamma^{\nu}+\Gamma^{\mu})^2/4}\label{EqAfalt}\, .
\end{equation}
This expression has a peak at the resonance $E^{\mu}=eFd+E^{\nu}$
and a full width at half maximum of $(\Gamma^{\nu}+\Gamma^{\mu})$.
Even if the conditions stated above are not fulfilled, the result
is typically similar.

For  tunnelling between the lowest levels $a\to a$, equal densities
$E^F_n=E^F_{n+1}$, a constant density of states  $\rho^a$
and not too high temperatures and fields  $k_BT,eFd<E^F_n-E^a$  one finds 
\begin{equation}
I_{n\to n+1}^{a\to a}=e A \rho^a \frac{(T_1^a)^2}{\hbar}
\frac{2\Gamma^a eFd}{(eFd)^2+(\Gamma^a)^2}\label{Eqatoa}\, .
\end{equation}
Therefore the current has a maximum at $eF_{\rm max} d=\Gamma^a$ where
it takes the value $I_{\rm max}=e A \rho^a (T_1^a)^2/\hbar$.
Note that this value neither depends on $\Gamma$ nor on the Fermi level.
Estimating $\Gamma\approx 8$ meV from the spectral functions shown
in Fig.~\ref{Figspektral} and taking $\rho^a=\rho_0$ yields
\begin{equation}
eF_{\rm max}d\approx 8{\rm meV} \qquad I_{\rm max}\approx 0.27 {\rm mA} \, .
\label{Eqestimate}
\end{equation}
for the $a\to a$ transition using $T_1^a=-0.0201$ meV from 
Table~\ref{Tabcoeff}.

With the translations $2T_1^a \to E_1$, $E^a\to E_0$, and 
$\hbar/\Gamma^a \to \tau$ 
Eq.~(\ref{Eqatoa}) is identical to equation~(12)
\footnote{Note that there is a factor $\tau$  missing in the numerator 
due to a misprint.} of Ref.~\cite{LEB70} for the case $E^F>E_0+E_1$. 
There semiclassical transport in a miniband for
the strong coupling limit $T_1^a\gg \Gamma^a$ was regarded
while here the limit of weakly coupled quantum wells
$T_1^a\ll \Gamma^a$ is considered. Note that in both derivations the case 
$(E^F-E^a)>\Gamma^a,2T_1^a$ is considered.

The transitions from the lowest level to the excited level  $a\to b$
will become important if $eFd\approx E^b-E^a$. Assuming that
$k_BT,E_{n+1}^F\ll E^b-E^a$ we find $n_F(E-E_{n+1}^F+eFd)\approx 0$
in this field range. Using the approximation (\ref{EqAfalt}) we find 
close to the resonance
\begin{equation}
I_{n\to n+1}^{a\to b}\approx    
e A n_n^a \frac{|H_{1}^{{\rm res}\, a,b}|^2}{\hbar}
\frac{\Gamma^{a}+\Gamma^{b}}
{(eFd+E^{a}-E^{b})^2+(\Gamma^{a}+\Gamma^{b})^2/4}\label{Eqatob}\, .
\end{equation}
Note that this current is proportional to the density of carriers
$n_n^a=\int dE n_F(E-E_n^F)\rho^{a}(E)$ in well $n$  while this was
not the case for the $a\to a$ peak in Eq.~(\ref{Eqatoa}).

\subsection{Nonresonant Current}
The transition elements from $\hat{H}^{{\rm scatter}}_1$ do not conserve 
momentum and therefore the kinetic energy $E_k$ is changed during the
transition. Thus these transitions are not  as sensitive to the alignment
of the levels as the resonant transition discussed above.
They yield a background current which may dominate the current
between the resonances.
Here only nonresonant transitions via interface roughness are considered.
The explicit expressions used are given in section~\ref{Secdetails}.
As the actual shape of the interface  is not known 
and may strongly vary  for different wafers
it has been parameterized by a  reasonable set of parameters in order
to show the magnitude of the effect.
Therefore the reader  has to keep in mind that  the nonresonant currents
can easily vary by a factor of three or even more in the 
following calculations.

\subsection{Calculation of the current}
\begin{figure}
\vspace*{5.7cm}
\includegraphics{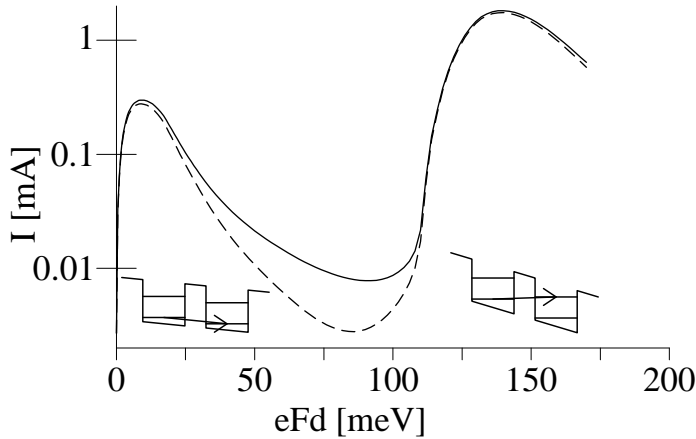}
\caption[a]{Current $I_{i\to i+1}$ for $n_i=n_{i+1}=N_D$.
The dashed line gives the current from resonant transition,
while the full line gives then sum of resonant and nonresonant
transitions due to interface roughness.} 
\label{Figstromhom}
\end{figure}

In Eq.~(\ref{EqJ}) the current depends on the electric field and on the
Fermi energies in both wells.
The Fermi energies $E_i^F$ can easily be calculated from 
the two-dimensional density of carriers $n_i$ in the well using the  relation
\begin{equation}
n_i=\int dE n_F(E-E_i^F)\sum_\mu \rho^{\mu}(E)\, .
\end{equation}
Finally, the total current from well $i$ to well $i+1$
is the sum of 
the contributions between the different levels.
\begin{equation}
I_{i\to i+1}=\sum_{\mu,\nu}I_{i \to i+1}^{\nu \to \mu}
=I(F,n_i,n_{i+1})
\label{EqItotal}
\end{equation}
The result is presented in Fig.~\ref{Figstromhom} for $n_i=n_{i+1}=N_D$.
Neglecting the nonresonant transitions (dashed line) there is
a first maximum at  $eFd=9$ meV with a current of $0.277$ mA
which is in good agreement with the estimation (\ref{Eqestimate})
for the $a\to a$ transitions.
A second maximum occurs at $eFd=140$ meV with $I=1.75$ mA.
In comparison to this the experimental data (see Fig. 6 of \cite{KWO95}) 
exhibit a 
first maximum of $I\approx 0.076$ mA and a second maximum of
$I\approx 1.45$ mA. While there is good quantitative agreement
for the height of second maximum, the calculated first maximum  
seems to be too large by a factor of 4. This inconsistency will be 
resolved in the next chapter where the formation of field domains 
is considered. In contrast to the currents, 
the position of the maxima is much more difficult to compare as
a part of the voltage may drop outside the superlattice.
The nonresonant currents are not very large compared to the
currents at the maxima but can dominate the total current between
the maxima as can be seen from the full line  in Fig.~\ref{Figstromhom}.

Note that the experimental data of Ref.~\cite{KWO95} indicate that
the first excited level is unoccupied close to the onset of
the $a\to b$ resonance indicating that the current of
approximately 0.1 mA is not carried by the $a\to b$ transitions there.
This may be attributed to stronger nonresonant transition in this 
field range.

An important feature of the model presented here is the
fact that only the nominal sample parameters are involved in the
calculation of the currents. The resulting currents are in good quantitative
agreement with the experimental data of Refs.~\cite{GRA91,KWO95}.
The same model has also been applied to the samples used in Ref.~\cite{HEL90}
where quantitative agreement could be obtained  assuming a smaller 
barrier width \cite{WACpa}. While these two highly-doped samples
exhibit a density of states which resembles the free-electron 
density of states with a smoothed 
onset (see Fig.~\ref{Figspektral}),
the density of states is much more complicated  for low-doped samples
due to the presence of impurity bands \cite{GOL88}. Within the single-site
approximation for impurity scattering  used here (see section \ref{Secdetails})
these effects are included in the spectral function.
The calculation yields a strong dependence of the current on temperature
in good agreement with experimental data \cite{WACpc}.
Excellent quantitative agreement has also been found when the  
sample was irradiated by a strong terahertz field from a free-electron 
laser source \cite{WACpc}.
Therefore we may conclude that the formalism described here allows the
quantitative calculation of the currents in weakly-coupled multiple 
quantum wells for a wide range 
of samples without applying any fitting parameters.

\section{Formation of field domains}

Now we want to consider the full structure consisting of $N$ wells
numbered $i=1,\ldots N$ and $N+1$ barriers.
Assuming that the electric field is constant over the structure
the total voltage is given by $U\approx (N+1)Fd$.
Fig.~\ref{Figstromhom} shows  that $I_{i\to i+1}(F,N_D,N_D)$ exhibits  
a region of negative differential conductivity (NDC)
between $eF_{\rm max}d=9.7$ meV and $eF_{\rm min}d=90$ meV. 
If the electric field is  within this NDC-region 
an instability  is likely to occur because a spontaneously formed 
charge accumulation will increase in time instead of
decreasing. This is a common situation for samples exhibiting an N-shaped
local current-field relation like the Gunn diode (see, e.g., \cite{SHA92}
and references cited therein).
In order to include such effects 
charge accumulation  has to be allowed for. 
This means that the
carrier density $n_i$ in well $i$ may deviate from the doping $N_D$
per period. Integrating Poisson's equation over one period yields
\begin{equation}
\epsilon_r\epsilon_0 (F_i-F_{i-1})=e(n_i-N_D)
\quad \mbox{for } i=1,\ldots N
\label{Eqpoisson}
\end{equation} 
where $F_i$ is the electric field in the middle of the barrier between
wells $i$ and $i+1$. Here a constant dielectric permeability
$\epsilon_r=13$ is assumed  for simplicity. 
\begin{figure}
\vspace*{4.92cm}
\includegraphics{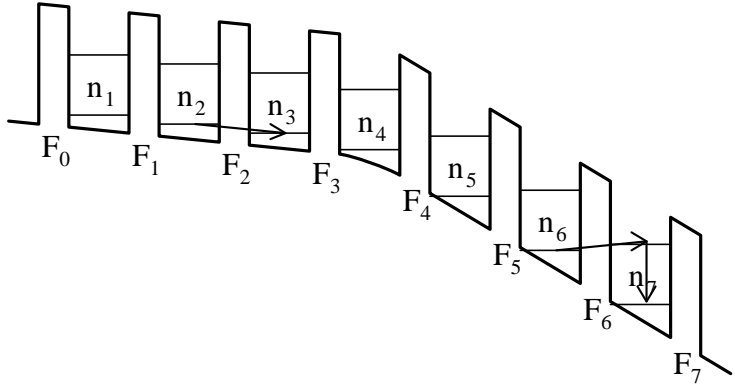}
\caption[a]{Sketch of the fields and densities for domain formation.} 
\label{Figskizze}
\end{figure}

The notation as well as a typical potential profile is 
sketched in Fig.~\ref{Figskizze} for a superlattice with $N=7$.
Within the first wells there is 
$n_1=n_2=n_3=N_D$ so that $eF_0=eF_1=eF_2=eF_3$.
The electric field is low so that the current is dominated
by the $a\to a$ transitions. At well 4 there 
is charge accumulation $n_4>N_D$. Therefore we have $eF_4>eF_3$
and the electric field is  large after well 4 so that the current
is dominated by the $a\to b$ resonance. Here we assume that the
electrons relax  fast from the upper levels into
the lower level which is the only level to be populated in thermal equilibrium
for level separations larger than the Fermi energy.\footnote{Furthermore
the  intersubband relaxation must be fast enough to guarantee
thermal equilibrium between the levels.
If the level separation is larger than the optical phonon energy 36  meV,
the intersubband relaxation time is of the order of 1 ps\cite{FER89}, 
which is typically
much faster than the tunnelling times. But for wide quantum wells
the level separation becomes small and nonequilibrium effects are 
found\cite{MITp}.}

In order to calculate the current between the wells the question arises
which electric field should be used if an inhomogeneous
situation is considered. As the voltage drop between the wells $i$ and
$i+1$ can be approximated by $eF_id$ it is natural to use the electric field
$F_i$ in the argument of Eq.~(\ref{EqItotal}). Nevertheless it has to
be stated that this is an approximation done for simplicity and deviations
may occur for inhomogeneous field profiles.

Now the temporal evolution of the densities within the wells is given by the 
continuity equation
\begin{equation}
eA\frac{dn_i}{dt}=I_{(i-1)\to i}-I_{i\to (i+1)}
=I(F_{i-1},n_{i-1},n_i)-I(F_{i},n_{i},n_{i+1})
\label{Eqcontinuity}
\end{equation}
for $i=1,\ldots N$.
In order to obtain a complete set of equations for the temporal
evolution of the densities and fields we have to 
add two more features. At first the voltage condition
is now given by
\begin{equation}
U=\int dz F(z) \approx \sum_{i=0}^N dF_i +U_c
\label{Eqvoltage}
\end{equation}
where $U_c$ represents the voltage drop outside the superlattice, 
which is neglected in the following in order to concentrate on the
features of the pure superlattice.
At second the currents $I_{0\to 1}$ and 
$I_{N\to (N+1)}$ across the first and the last barrier
of the structure, respectively, have to be specified.
For simplicity one may  use the expression (\ref{EqItotal})
with  appropriate effective densities 
\begin{equation}
n_0=n_0(F_0,N_D,n_1)\quad\mbox{and} \quad n_{N+1}=n_{N+1}(F_N,N_D,n_N)\, .
\label{Eqboundary}
\end{equation}
In mathematical terms the  
functions $n_0(F_0,N_D,n_1)$ and $n_{N+1}(F_N,N_D,n_N)$ then 
represent the boundary
conditions of the model.
Eqs.~(\ref{Eqpoisson}-\ref{Eqboundary}) form a complete set of equations
for calculating the densities, fields, and currents as a function of time
for fixed bias voltage $U$ and
given initial conditions $n_i(t_0)$ (i=1,\ldots N).   

Using a different approach for evaluating the current function 
$I(F_{i},n_{i},n_{i+1})$
(which is essentially based on the broadening due to the tunnelling time
for the $a\to b$ resonance
and miniband conduction for the $a\to a$ resonance)
and the boundary conditions
$n_0(F_0,N_D,n_1)=n_1$, $n_{N+1}(F_N,N_D,n_N)=n_N$ such a model
has been used in Ref.~\cite{PRE94} to calculate the
current-voltage characteristics under domain formation.
The model of Ref.~\cite{BON94} uses the boundary condition
$n_0(F_0,N_D,n_1)=N_D$ and the current
expression 
\begin{equation}
I_{i\to (i+1)}=eAn_iv(F_{i})\label{Eqbonsimp}
\end{equation}
where $v(F)$ is a phenomenological tunnelling rate yielding
an N-shaped characteristic as shown in Fig.~\ref{Figstromhom}.
This simplification allows the analytical construction of the
full current-voltage characteristic \cite{BON94,WAC97a}.
Eq.~(\ref{Eqbonsimp}) is motivated
by Eq.~(\ref{Eqatob}) for the $a\to b$ transitions and higher resonances
as well. Nevertheless, is seems to be questionable close to
the $a\to a$ maximum as the maximum current of Eq.~(\ref{Eqatoa})
is independent of $n_i$ for equal carrier densities in adjacent wells.
The reason for this deviation is the existence of backward currents
for low fields which are taken into account by the factor 
$\left[n_F(E-E_n^F)-n_F(E-E_{n+1}^F+eFd)\right]$ in Eq.~(\ref{EqJ}).
If on the other hand $eFd\gg E_{n+1}^F-E^a+k_BT_e$ 
so that $n_F(E-E_{n+1}^F+eFd)\approx 0$ for all relevant energies $E$, 
Eq.~(\ref{Eqbonsimp}) is a reasonable approximation.

\subsection{Numerical  results}
\begin{figure}
\vspace*{8,6cm}
\includegraphics{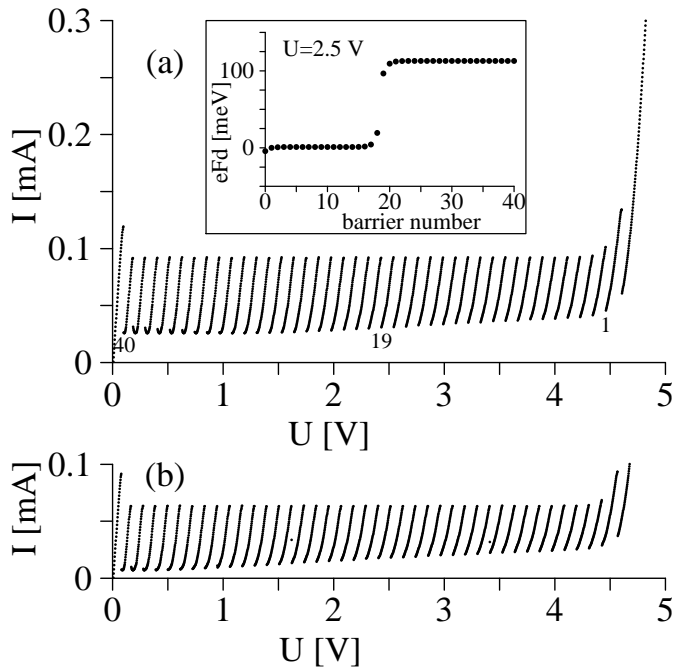}
\caption[a]{Calculated current-voltage characteristic exhibiting field
domains for voltage sweep-up for the boundary conditions
$n_0=3N_D$ and $n_{N+1}=3N_D$ with (a) and without (b)
nonresonant transitions due to interface roughness.
The inset gives the field distribution at $U=2.5$ V where the domain 
boundary is located at the 19th well.} 
\label{Figgrahndom3}
\end{figure}

In a first simulation the boundary conditions
$n_0(F,N_D,n_1)=n_{N+1}(F,N_D,n_N)=3N_D$ are used.
The stationary stable  states are determined
by simulating Eqs.~(\ref{Eqpoisson}-\ref{Eqboundary}) until a stationary
state is reached for given $U$. Increasing or decreasing $U$ afterwards
simulates a sweep-up or sweep-down of the voltage, respectively.
The result is shown in Fig.~\ref{Figgrahndom3}(a).
We find that the characteristic consists of 40 branches, equal to
the number of quantum wells. 
The maximum current within the branches is about 0.09 mA in good agreement
with the experimental situation 
\cite{GRA91,KAS94} where around 0.06 mA is observed.
This current is significantly lower than the first maximum
of  $I(F,N_D,N_D)$ (see Fig.~\ref{Figstromhom}) resolving the discrepancy 
discussed at the end of the last section.
Note that another
superlattice with the same specification \cite{WAC95c} 
exhibits maximum currents of the branches of 0.14 mA. 
This indicates that the current is quite sensitive
to variations in the sample.
The calculation exhibits similar effects. If we ignore the nonresonant 
transitions the maxima of the branches are found at 0.06 mA
as shown in Fig.~\ref{Figgrahndom3}(b), although the height of the maxima 
in the homogeneous characteristics (see Fig.~\ref{Figstromhom})
is almost identical.
Thus the extension of the branches
are very sensitive to the quality of the interfaces (which causes  
nonresonant transition via  interface roughness). 
Another uncertainty
may be the actual barrier thickness as a variation of one monolayer
changes the matrix elements $\hat{H}_1$ by a factor of 1.4 and therefore
the current (which is proportional $H_{i,i+1}^2$) by a factor of 2.

The slope of the branches varies between 1.1 mS and  0.55 mS 
for low and high voltage in  Fig.~\ref{Figgrahndom3} which 
is significantly larger than the experimental slopes 
(see \cite{KAS94}) varying between 0.133 mS and 0.064 mS.
This lower slope is responsible for the stronger overlap between
the branches yielding pronounced multistability \cite{KAS94}. 
The discrepancy may be resolved by assuming 
an appropriate contact voltage $U_c$ in Eq.~(\ref{Eqvoltage})
which will depend on $I$ and the fields at the boundaries.
Another possibility might be that further nonresonant transitions
alter the shape of the homogeneous characteristics of Fig.~\ref{Figstromhom}.

The field distribution shown in the inset of Fig.~\ref{Figgrahndom3}(a)
is in good agreement with  cathodoluminescence
measurements  \cite{KWO95b} stating that the field distribution
consists of one low field and one high field domain where the 
high field domain is located at the anode. These measurements where
performed at the same sample but in the range between the $a\to b$
and the $a \to c$ resonance.

\begin{figure}
\vspace*{7.4cm}
\includegraphics{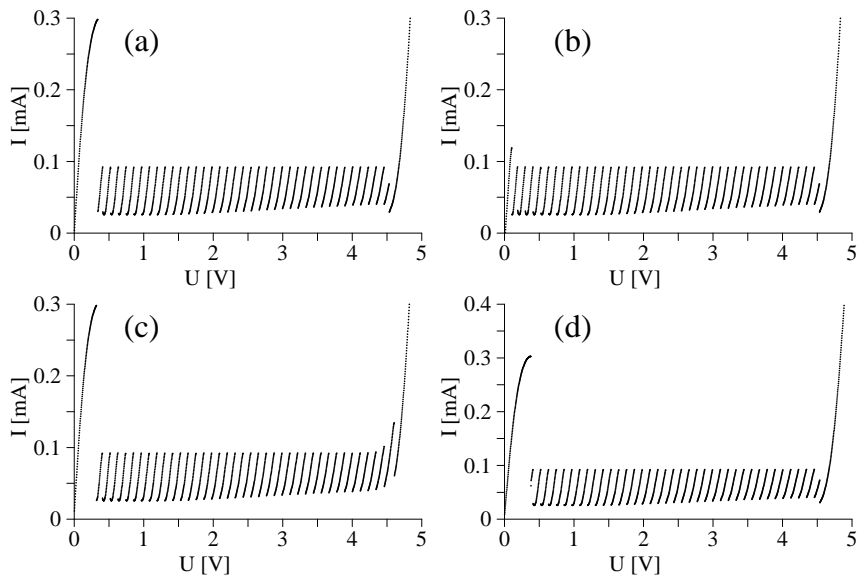}
\caption[a]{Calculated current-voltage characteristic 
for voltage sweep-up for the boundary conditions
$n_0=1.1N_D$, $n_{N+1}=1.1N_D$ (a),
$n_0=1.1N_D$, $n_{N+1}=3N_D$ (b),
$n_0=3N_D$, $n_{N+1}=1.1N_D$ (c), and
$n_0=1.2N_D$, $n_{N+1}=0.8N_D$ (d)} 
\label{Figgrahndomdiv}
\end{figure}

Using different boundary conditions   almost identical
domain branches are observed as shown in Fig.~\ref{Figgrahndomdiv}. 
The main difference occurs at the
first and last branch which may be changed significantly.
Comparing Fig.~\ref{Figgrahndomdiv}(a,b,c) 
and Fig.~\ref{Figgrahndom3}(a) shows that  the first branch is 
dominated by the boundary condition
$n_{N+1}(F_0,N_D,n_1)$ and the last branch by $n_0(F_N,N_D,n_N)$.

\subsection{General aspects of domain formation}
Now a qualitative explanation of the domain formation shall be given.
First regard Fig.~\ref{Figstromhom}. 
For  currents $I_0$ from the interval
$[I_{\rm min},I_{\rm max}]$ where 
$I_{\rm max}= 0.30$ mA is the maximum current for low fields
and $I_{\rm min}= 7.8\mu$A is the minimum current for medium fields
there are 3 intersections $F_{I}<F_{II}<F_{III}$ 
of the $I(eFd)$ curve with $I_0$. Unlike $F_{II}$  
the fields $F_{I}$ and $F_{III}$
are in the range of positive differential conductivity (PDC).
Therefore charge fluctuations are damped out in spatial regions where
the electric field takes the values $F_{I}$ or $F_{III}$.
The field distribution for $U=2.5$ V in the inset of 
Fig.~\ref{Figgrahndom3}(a) now shows that the 
electric field takes the value $F_i\approx F_{I}$ for $i\le17$ and
$F_i\approx F_{III}$ for $i\ge 20$.
The small deviations for $i=0,1$ due to the contact are damped
out because the field is in the PDC region there.
In between there is a domain boundary where the electric field
changes due to a charge accumulation $e(n_i-N_D)=\epsilon_r\epsilon_0
(F_{i}-F_{i-1})$. 
As the main jump occurs between $F_{18}$ and $F_{19}$
the density $n_i$ takes its maximum at $i_D=19$.
If now this charge accumulation layer is shifted by one period
to $i_D=20$, we have almost the identical situation
with the same current except for the fact that the voltage is
diminished by $(F_{III}-F_I)d$ as a lower fraction of the sample
is in the high field region.
This reveals the periodic sequence of branches.
If one counts the branches starting from the right side
as depicted in Fig.~\ref{Figgrahndom3}(a), the
maximum of $n_i$ occurs in well $i_D$ for the $i_D$th branch.
In total one can count 40 branches which is exactly the number of 
quantum wells as the domain boundary may be located in each well.
If the domain boundary comes close to
$i=1$ the low field region is not large enough to shield
the variation due to the  contact. Therefore the first and the second branch
are strongly dependent on the boundary condition which simulates
the contact (compare Fig.~\ref{Figgrahndom3}(a) with \ref{Figgrahndomdiv}).
The field distribution in such stationary domain structures is not arbitrary
as the currents across each barrier have to be equal.
This provides a condition on  the minimal doping density (or minimal carrier 
generation due to irradiation) as discussed in 
Refs.~\cite{BON94,SCH96b,WAC97a,MIT96}.

Now the question arises why these stationary domain states are stable
while in other spatially extended  NDC systems like the Gunn diode 
typically travelling field domains occur.
In Ref.~\cite{WAC97a} this question has been investigated using the
simplified current relation (\ref{Eqbonsimp}).
There it could be strictly proven that  an inhomogeneous field 
distribution is necessarily stable
if all electric fields $F_i$ are in the PDC region of
$v(F)$ which coincides with most readers' physical intuition.
Therefore a field distribution like that in the inset of 
Fig.~\ref{Figgrahndom3}(a) must be stable if the NDC region 
$eF_{\rm max}<eF<eF_{\rm min}$
is crossed within one jump. Then
\begin{equation} 
eF_{i_D-1}\le eF_{\rm max} \quad \mbox{and}\quad 
eF_{i_D}\ge eF_{\rm min}\label{Eqcondstab}
\end{equation} 
holds. In the stationary state the current across each barrier has  
to be equal to $I_0$. Especially, there is
$I_0=I(F_{i_D},n_{i_D},n_{i_D+1})$.
As $F_{i_D}$ is in the high field region  the approximation (\ref{Eqbonsimp})
may be justified yielding
$ I(F_{i_D},n_{i_D},n_{i_D+1})\approx I(F_{i_D},N_D,N_D)n_{i_D}/N_D$.
Combining this with Poisson's equation
$e(n_{i_D}-N_D)=\epsilon_r\epsilon_0(F_{i_D}-F_{i_D-1})$
exhibits that the condition (\ref{Eqcondstab}) can be fulfilled if
\begin{equation}
I_0>I(F_{\rm min},N_D,N_D)
\left(1+\frac{\epsilon_r\epsilon_0(F_{\rm min}-F_{\rm max})}{eN_D}\right)
\end{equation}
holds. It has to be stated that this is only a sufficient condition for 
stability as there are stable domain states where one field
is located within the NDC region. 
Now $I_0$ can not be larger than $I(F_{\rm max},N_D,N_D)$ 
as the low field domain can not carry a larger current.
Therefore there is a minimum doping density 
\begin{equation}
N_{\rm crit}^{\rm acc}\sim
\frac{\epsilon_r\epsilon_0(F_{\rm min}-F_{\rm max})}{e}
\frac{I_{\rm min}}{I_{\rm max}-I_{\rm min}}
\label{Eqacccondition}
\end{equation}
above which stable domain states with an accumulation layer exist.
From the $I(F_{i},N_D,N_D)$ relation from Fig.~\ref{Figstromhom}
$N_{\rm crit}^{acc}\approx 1.2\times 10^{10}/{\rm cm}^2$ is estimated
which is much smaller than $N_D=1.5\times 10^{11}/{\rm cm}^2$. Therefore
stable domain states are expected in accordance with the experimental
and theoretical findings. Note that the condition (\ref{Eqacccondition})
depends strongly on $I_{\rm min}$ which itself is strongly affected by
nonresonant transitions as shown in Fig.~\ref{Figstromhom}.

\begin{figure}
\vspace*{5.9cm}
\includegraphics{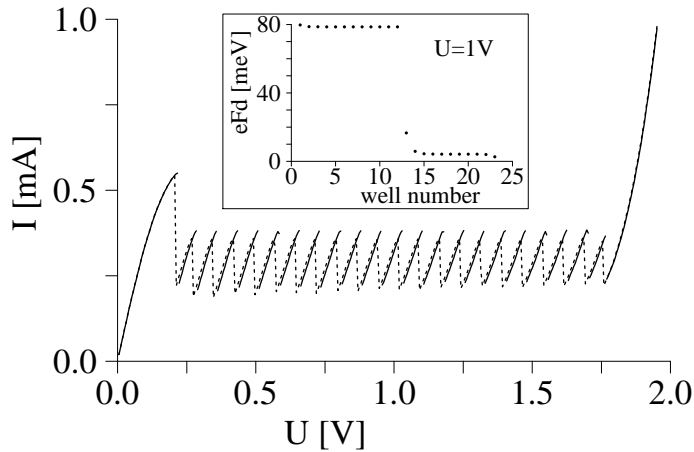}
\caption[a]{Calculated current-voltage characteristic 
of the sample from Ref.~\cite{HEL90} for
voltage sweep-up for the boundary conditions
$n_0=n_{N+1}=0.95N_D$ (full line),
$n_0=n_{N+1}=1.05N_D$ (dotted line)} 
\label{Fighelgedom}
\end{figure}

Up to now domain structures have been discussed where the high field
domain is located at the receiving contact.
But of course there is the other possibility that the high field
domain is at the injecting contact, i.e., 
$F_i\approx F_{III}$ for $i<i_D$ and $F_i\approx F_{I}$ for $i>i_D$.
Then Poisson's equation yields a depletion region $n_i<N_D$ at the domain
boundary $i\approx i_D$.  
Such domains have been observed experimentally for highly doped samples
\cite{HEL90}. Theoretically, such domains could be both obtained from
a simplified model \cite{WAC97a} and from the microscopic model \cite{WACpa}.
This is shown in Fig.~\ref{Fighelgedom}
using the parameters of the sample from Ref.~\cite{HEL90}.
For the boundary conditions
$n_0(F_0,N_D,n_1)=0.95N_D$, $n_{N+1}(F_N,N_D,n_N)=0.95N_D$
domain states are found, where the high field domain is located
at the injecting contact as shown in the inset, while
for  $n_0(F_0,N_D,n_1)=1.05N_D$, $n_{N+1}(F_N,N_D,n_N)=1.05N_D$
the high field domain is located
at the receiving contact like in Fig.~\ref{Figgrahndom3}.
The domain branches themselves look very similar.

Regarding the stability of such domain structures the 
argument given above yields a minimum doping 
\begin{equation}
N_{\rm crit}^{\rm dep}\sim
\frac{\epsilon_r\epsilon_0(F_{\rm min}-F_{\rm max})}{e}
\frac{I_{\rm max}}{I_{\rm max}-I_{\rm min}}
\end{equation}
above which stable domain states with a depletion layer can exist.
For the sample of Ref.~\cite{GRA91}  the estimation gives
$N_{\rm crit}^{dep}\approx 4.4\times 10^{11}/{\rm cm}^2$
which is three times larger than $N_D$.
Thus, such domain states are not expected to be stable and therefore
should not be observed.
For comparison the sample used in \cite{HEL90} has
$N_D=8.75\times 10^{11}/{\rm cm}^2$ and from Fig.~3 of Ref.~\cite{WACpa}
$N_{\rm crit}^{dep}\approx 2.5\times 10^{11}/{\rm cm}^2$ is estimated
in good agreement with the observation of stable domain structures
with depletion layers.

Note that the proof of stability essentially relies on the 
discreteness of the system.  In a continuous model
the NDC-region can not be crossed without having any fields
within this region at least for a small spatial interval.
This explains the difference to continuous systems like the Gunn diode
where such stable domain states with an arbitrary position of 
the boundary are not observed.

\section{Imperfect Superlattices}
\begin{figure}
\vspace*{4.0cm}
\includegraphics{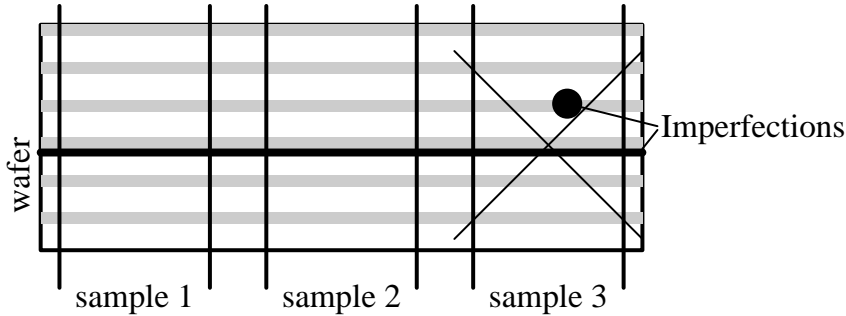}
\caption[a]{Schematic draw of a superlattice wafer from which 3 
different samples are obtained.
The dot in sample 3 represents a local imperfection while the line
represents a modulation of the periodic structure.} 
\label{Figwafer}
\end{figure}

All the theoretical current-voltage characteristics shown up to now
exhibited an almost regular series of branches whereas
in typical experiments the lengths of the branches vary.
It is straightforward to assume that  this is caused by irregularities
in the real superlattice as nothing is perfect.
But then the question about the nature of these irregularities arises.
At first there are two different possibilities:
Regarding a wafer as sketched in Fig.~\ref{Figwafer} the irregularity 
may either be a bad spot localized somewhere in the superlattice
as shown in the right side of the wafer or a deviation from
periodicity occurring in the whole layer. 
Such deviations may be
a larger or smaller barrier width, or a different doping density
in certain wells, e.g., which are established due to unsufficient control
of the growth process.
Fortunately, one can  distinguish these two cases experimentally.
If the local spot would be the essential cause for the irregularities
in the characteristics, sample 2 and sample 3 from the same wafer 
sketched in Fig.~\ref{Figwafer} should
exhibit a different modulation of the branches as their individual spots
have different sizes and are located at different positions.
But the different samples used in \cite{WAC95c} exhibited
almost identical modulations of the branches if they originated from the
same  wafers while there are large differences for samples
from different wafers even if the superlattice structure
is nominally identical. This would be expected from the samples
1 and 2 sketched in Fig.~\ref{Figwafer}.
The same feature can be observed in \cite{HEL90} where the authors
have fabricated  samples with different numbers of periods 
originating  from one wafer by an etching process.
They found very similar  sequences of longer and shorter branches
which allowed them to conclude that the high field domain is located at
the receiving contact in their samples.

These experimental observations clearly indicate that the dominant
deviations from periodicity are not $(x,y)$-dependent but extend over the
whole wafer. 
This allows to simulate these irregular superlattices 
by introducing local fluctuations in the well width, barrier width or
doping concentration into Eqs.~(\ref{EqJ},\ref{Eqpoisson}).
This has been done in Ref.~\cite{WAC95c} within the model of
Ref.~\cite{PRE94} where we found that even small spatial 
fluctuations (about 7\%) of the doping have a significant influence on 
the length of the branches and can explain the observed behaviour.
Furthermore, single fluctuations may be located by just determining
the number of the  branch which is altered most.
This effect is particularly pronounced for fluctuations in the 
barrier width, where a one monolayer fluctuation may change the characteristic
significantly.
The theoretical prediction has been successfully checked experimentally
by growing a new sample exhibiting one barrier with a larger
thickness \cite{SCH96d}. 
\begin{figure}
\vspace*{11.8cm}
\includegraphics{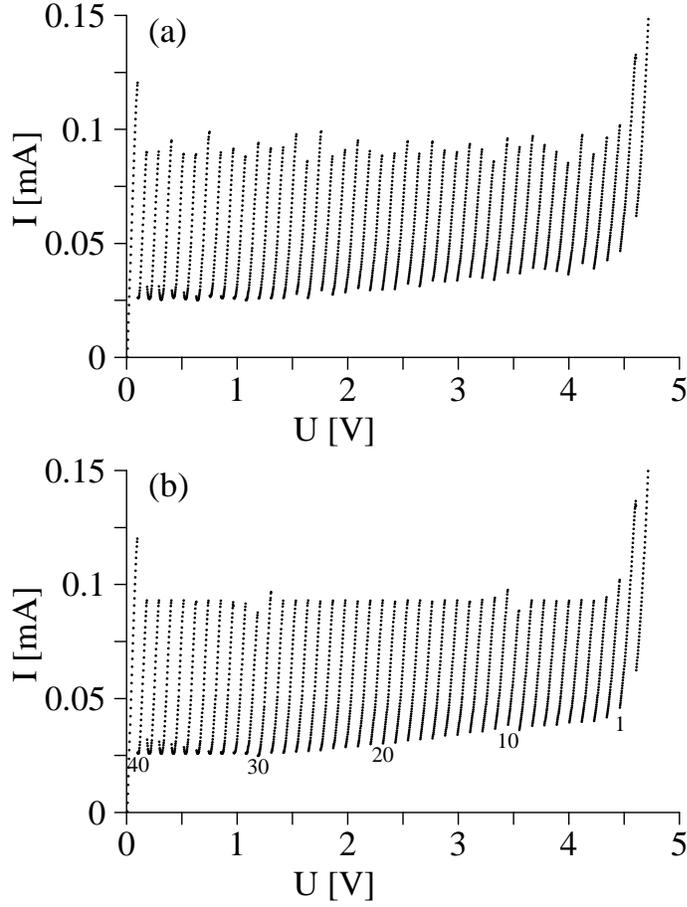}
\caption[a]{Current-voltage characteristic for doping fluctuations,
In (a) a random sequence $N_{Di}$ with an average fluctuation
of 10\% around $N_D$ is used. In (b) there is $N_{D10}=1.2N_D$ and
$N_{D30}=0.8N_D$ while the other densities are not altered.} 
\label{Figdotfluk}
\end{figure}

Now doping fluctuations are included into the model discussed before
by using a well-dependent doping density $N_{Di}$
in Eq.~(\ref{Eqpoisson}). The impact on the impurity scattering is neglected,
so that the old function $I(F_i,n_i,n_{i+1})$ is used.
The result is shown in Fig.~\ref{Figdotfluk} and exhibits
fluctuating branch heights in accordance with the model used in 
Ref.~\cite{WAC95c}. Comparing the local fluctuations $N_{Di}$ with the
maximum current of the branch, one finds a  correlation.
Like the findings of Ref.~\cite{PAT96} the branch $i+1$ counted from
the right extends to higher currents if  $N_{Di}$ is larger than $N_D$.
Additionally the branch $i$ extends to lower currents as can be seen 
from Fig.~\ref{Figdotfluk}(b). The deviations in the current seem 
to be smaller and the effects to neighboured branches seem to be larger here
than observed in Refs.~\cite{WAC95c,PAT96}. This might be related to
the fact that a higher doping density was used in previous works.

\section{Oscillatory behaviour}
\begin{figure}
\vspace*{11.8cm}
\includegraphics{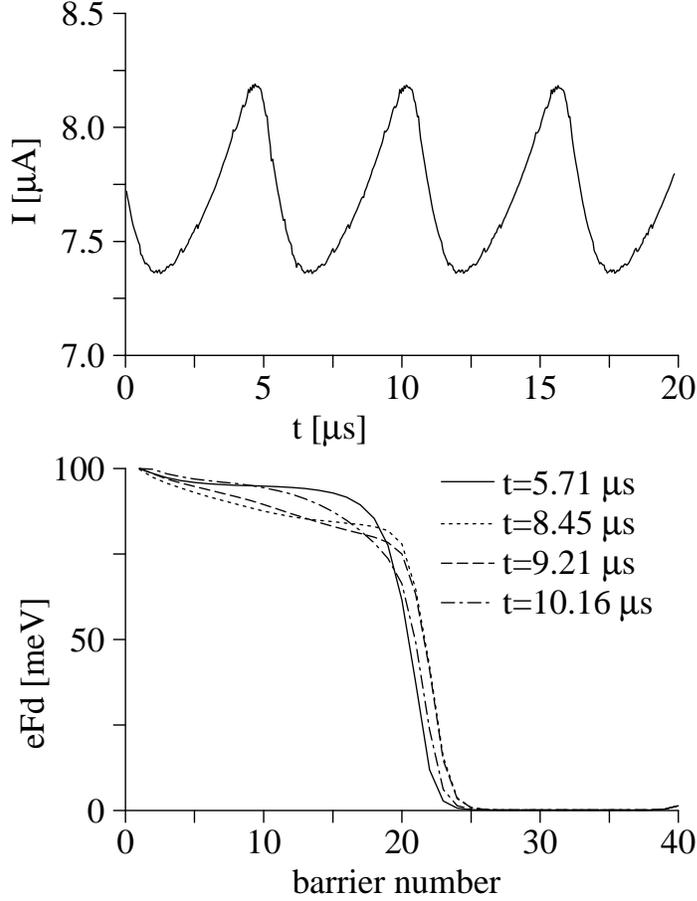}
\caption[a]{Calculated current oscillations for $U=2$ V.} 
\label{Figgrahnosc}
\end{figure}

Oscillatory behaviour has been observed in 
coupled multiple quantum wells both experimentally\cite{KAS95,GRA96,OHT96}
and theoretically \cite{BON95,WAC95}.
As an example such behaviour is obtained within the model discussed
here for the boundary conditions 
$n_{0}(F,N_D,n_1)=0.8 N_D$ and  $n_{N+1}(F,N_D,n_N)=1.2 N_D$ which is just the
reversed sequence used for the calculation of Fig.~\ref{Figgrahndomdiv}(d)
where stable domain branches were found.
Fig.~\ref{Figgrahnosc} shows self-sustained current oscillations
between 7.4 $\mu$A and 8.2 $\mu$A  with a frequency of 0.18 MHz. 
Similar results are  obtained  for different biases and different 
values $n_0(F,N_D,n_1)< N_D$.
The oscillations may be described from the field distribution
as follows.
At $t=5.71\mu$s the field distribution consists of a high field and a low
field domain
with a depletion layer in between. As for $N_D<N^{\rm dep}_{\rm crit}$
such a distribution is not stable
the boundary travels to the right thereby decreasing the electric field
in the high field domain because the total voltage has to remain
constant ($t=8.45\mu$s).
As the high field domain enters now the region of NDC for $eFd<90$ meV 
a new positive charge accumulation  layer is created there which is 
slightly visible for $t=9.21\mu$s at $i\approx 10$.
This accumulation layer travels to the right and increases in time
($t=10.16\mu$s) until it merges with the old depletion layer.
Then the cycle is repeated again. Such an oscillation type has
been described  in Ref.~\cite{KAS97} within the approximation
(\ref{Eqbonsimp}).
The very same behaviour happens for charge accumulation layers as well
if domain states with an accumulation layer are unstable.
For a full treatment of this oscillation type see Ref.~\cite{BON97}.
Note that the minimum current in this oscillation cycle 
roughly coincides with $I(F_{\rm min},N_D,N_D)$.
This relation seems to hold generally as checked by altering
the nonresonant current  and thereby $I(F_{\rm min},N_D,N_D)$.
Therefore the temporal minimum of this type of current oscillation
provides information about $I(F_{\rm min},N_D,N_D)$ which is strongly
dependent on the nonresonant transitions.

\begin{figure}
\vspace*{11.8cm}
\includegraphics{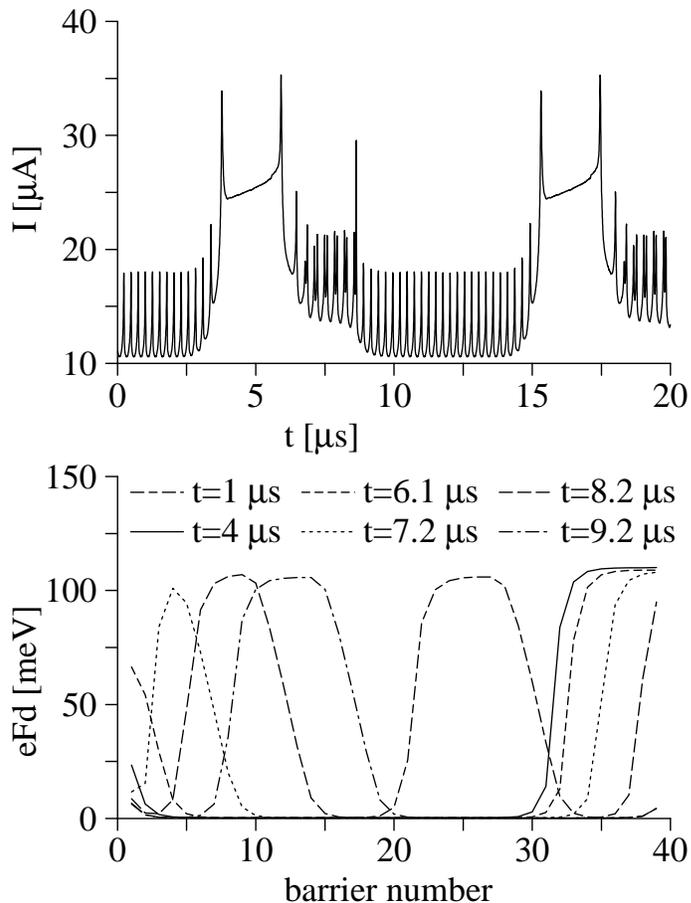}
\caption[a]{Calculated current oscillations for $U=1$ V using ohmic 
boundary conditions with $\sigma=0.5$ mA/eV.} 
\label{Figohmosc}
\end{figure}

In order to study the influence of the boundary conditions onto 
the oscillations we use the ohmic 
contact currents $I_{0\to 1}=\sigma eF_0d$ and  
$I_{N\to N+1}=\sigma eF_Nd$ in the following.
For $\sigma=0.5$ mA/eV a completely different
oscillation mode is found as shown in Fig.~\ref{Figohmosc}.
Here one oscillation cycle consists of the nucleation
and travelling of a dipole domain which vanishes by leaving the sample
at the receiving contact. 
The scenario is completely analogous to that described in \cite{BONp}
for a continuous system exhibiting NNDC.
Additionally, small current spikes appear which are related to
the motion of the accumulation layer  from one well to the next 
as discussed in Ref.~\cite{KAS96} for the domain formation process.
Thus, these spikes reflect the discreteness of the superlattice.

Experimentally, the sample discussed here also exhibits 
self-sustained current oscillations
if a positive bias is applied at the top contact while there are stable
domain branches for a negative bias.
The experimental data \cite{KAS97} exhibit current oscillations
between 20 $\mu$A and  35 $\mu$A with a frequency of 0.4 MHz
at $U=0.78$ V. Thus, the theory is in qualitative agreement with 
the experiment regarding the oscillations as well.
As we did not try and  model the contact currents 
microscopically (which strongly influence the type
of oscillations) we can not expect quantitative agreement.
Additionally, the shape of the oscillations as well as the frequency
depends substantially on the full shape of the current-field relation
which is strongly affected by the nonresonant transitions between
the resonances. Furthermore, deviations from periodicity may affect
the oscillations as well \cite{WAC95c,SCH96}.

For larger values of $\sigma$ (say $\sigma\gtrsim 3$ mA/eV) the same 
domain branches like those shown in  Fig.~\ref{Figgrahndomdiv} are found. 
This indicates that the domain branches
themselves are almost identical if their formation is allowed for
by the contact conditions and if the doping is sufficiently large.
On the other hand the oscillatory behaviour does strongly depend
on the exact contact conditions.

\section{Details of the calculations \label{Secdetails}}
In this last section I want to present the  details of the 
calculations whose results have been shown before.

\subsection{Calculation of the couplings}
In order to calculate the coefficients of Table~\ref{Tabcoeff} for  
Eqs.~(\ref{Eqham1}-\ref{Eqham3}) we have to specify the 
band structure in GaAs and AlAs  at first.
It is assumed that only the $\Gamma$ band is of importance.
While one may use a parabolic band structure for Al$_x$Ga$_{1-x}$As/GaAs
heterostructures for small $x$ this is not appropriate for an 
GaAs/AlAs heterostructure as the conduction band of GaAs is located far
in the band gap of AlAs where the band structure is
clearly not parabolic (see Ref.~\cite{SCH85a}).
Following Ref.~\cite{BRO90} we model the nonparabolicity
by an energy dependent effective mass $m(E)=m_c(1+(E-E_c)/E_g)$,
where $m_c$ is the effective mass at the conduction band minimum of 
energy $E_c$, and $E_g$ is the energy gap. Then the usual connection rules
\begin{eqnarray}
\varphi(z_0-\varepsilon)&=& \varphi(z_0+\varepsilon)\\
\frac{1}{m(E,z_0-\varepsilon)}\frac{{\rm d}\varphi}{{\rm d}z}(z_0-\varepsilon)
&=&
\frac{1}{m(E,z_0+\varepsilon)}\frac{{\rm d}\varphi}{{\rm d}z}(z_0+\varepsilon)
\end{eqnarray}
hold for the envelope function $\varphi(z)$ 
provided that the momentum matrix element $P=\hbar\sqrt{E_g/(2m_c)}$
between the conduction and valence band states is identical in 
both materials. We use the values \cite{BRO90} $m_c^{{\rm GaAs}}=0.067 m_e$,
$m_c^{{\rm AlAs}}=0.15 m_e$, $E_g^{{\rm GaAs}}=1.52$ eV, 
$E_g^{{\rm AlAs}}=3.13$ eV, and the conduction band 
discontinuity $\Delta E_c=1.06$ eV.
These parameters yield a relation $E(k)=E_c+\hbar^2k^2/(2m(E))$ 
which is in excellent agreement with the band structure of 
AlAs\cite{SCH85a} for the energies of interest.
It must be stated that for these parameters  the value of 
$P$ is slightly different for the two materials in contrast to 
the assumption. 
Nevertheless, these parameters seem to give reasonable 
agreement with experimental data both in Ref.~\cite{BRO90}
and the calculations presented here.

Imposing the Bloch condition $\varphi_q^{\nu}(z+d)=e^{iqd}\varphi_q^{\nu}(z)$
the Bloch functions $\varphi_q^{\nu}(z)$ and eigenvalues $E_q^{\nu}$
are calculated within the Kronig-Penney model. The phase of the Bloch
functions is chosen in the following way \cite{KOH59}:
Let $z=0$ be the center of one quantum well.
If $\varphi_0^{\nu}(0)\neq 0$ we chose the phase in such a way that
$\varphi_q^{\nu}(0)$ is real for each $q$.
For $\varphi_0^{\nu}(0)=0$, $\varphi_q^{\nu}(0)$ is chosen to
be purely imaginary. Furthermore $\varphi_q^{\nu}(z)$ has to be an analytic 
function in  $q$ for both cases.
From Eq.~(\ref{Eqfourier}) the level energies $E^{\nu}$ and couplings
$T_1^{\nu}$ are obtained. Eq.~(\ref{Eqwannier}) provides the Wannier functions
which are plotted in Fig.~\ref{Figwannier}.
Finally, the  couplings $R_h^{\nu'\nu}$ are obtained from their definitions
$R_h^{\nu'\nu}=\int dz \Psi^{\nu'}(z-hd)z\Psi^{\nu}(z)$.
The calculated values are given in Table~\ref{Tabcoeff}.
A complication arises due to the fact that the effective Hamiltonian
of the  Kronig-Penney model
is energy dependent due to the energy dependence of the effective mass.
Therefore the envelope functions $\varphi_q^{\nu}(z)$ 
for different energies $E_q^{\nu}$ are not strictly orthogonal 
but exhibit a small overlap which is neglected in the calculation.

\subsection{Impurity scattering}
Here we want to calculate the self-energy\footnote{For readers who are not
familiar with the concepts of many particle physics 
(such as self-energies, Green functions, etc.) Ref.~\cite{MAT92} 
is recommended as a helpful introduction.}
for impurity scattering.
The contribution to $\hat{H}^{{\rm scatter}}_0$ of 
Eq.~(\ref{EqH0scatter}) for the lowest level $a$ is given by
\begin{equation}
\hat{H}_{0}^{\rm imp}=\frac{1}{A}\sum_{\ul{k},\ul{p},i,n}
V_{i,n}^{aa}(\ul{p})a_n^{\dag}(\ul{k}+\ul{p})a_n(\ul{k})\label{EqH0imp}
\end{equation}
Here the subscript $i$ denotes the impurity
located at the position  $(\ul{r}_i,z_i)$. 
The matrix element is calculated with the Wannier functions yielding:
\begin{eqnarray}
V_{i,n}^{aa}(\ul{p})&=&\int d^2r\,dz\,e^{-i\ul{p}\cdot\ul{r}}
\Psi^*_a(z)\Psi_a(z)\frac{-e^2}
{4\pi\epsilon_s\epsilon_0\sqrt{|\ul{r}-\ul{r}_i|^2+(z-z_i+nd)^2}}\nonumber \\
&=&\frac{-e^2}{2\epsilon_s\epsilon_0 p}\int dz\,\Psi^*_a(z)\Psi_a(z)
e^{-p|z-z_i+nd|}
e^{-i\ul{p}\cdot\ul{r_i}}\, .
\end{eqnarray}
Note that the contribution for $p=0$ is canceled by the respective part
in the electron-electron interaction
as usual if the number of donors is equal to the number of carriers in
the whole sample.
In order to prevent  the divergence of
the matrix element for $p\to 0$ screening due to the electron-electron 
interaction has to be considered. (This is a general problem 
for the Coulomb interaction, see also chapters 5 and 6.)
The  Hamiltonian for the electron-electron interaction  reads:
\begin{eqnarray}
\hat{H}^{ee}&=&\frac{1}{2A}\sum_{\ul{k},\ul{k'},\ul{p},n,h}
W_{h}^{aaaa}(\ul{p})a_n^{+}(\ul{k}+\ul{p})a_{n+h}^{+}(\ul{k'}-\ul{p})
a_{n+h}(\ul{k'})a_n(\ul{k})\nonumber \\
&&\qquad + \mbox{terms with b}\, .
\end{eqnarray}
Here we assumed that the overlap of wave functions from different
wells is negligible, so that pairs always have to be inside the same well.
The matrix element reads:
\begin{eqnarray}
W_{h}^{\mu\mu'\nu'\nu}(\ul{p})=
\frac{e^2}{2\epsilon_s\epsilon_0 p}\int dz_1\int dz_2\,
\Psi^{\mu*}(z_1)\Psi^{\mu'*}(z_2-dh)\nonumber\\
\times \Psi^{\nu'}(z_2-dh)\Psi^{\nu}(z_1)e^{-p|z_1-z_2|}
\end{eqnarray}

\begin{figure}
\vspace*{4.5cm}
\includegraphics{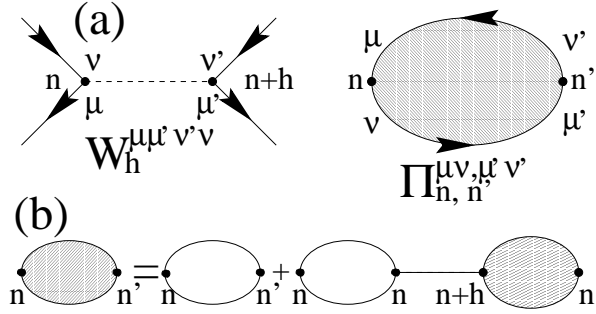}
\caption[a]{a) gives the notation of the matrix element 
$W_{h}^{\mu\mu'\nu'\nu}(\ul{p})$ and the 
polarizability $\Pi^{\mu\nu,\mu'\nu'}_{n,n'}(\ul{p},\omega=0)$.
b) shows the random-phase approximation diagramatically}
\label{Figscreen}
\end{figure}

The screening is described by the polarizability
$\Pi^{\mu\nu,\mu'\nu'}_{n,n'}(\ul{p},\omega=0)$, 
where $\nu,\mu$ take the values $a$ and $b$.
The notation follows Ref.~\cite{ELI87}, where a similar problem
is investigated, and
is shown in Fig.~\ref{Figscreen}. 
As the $\ul{p}$
dependence is identical in all parts, we omit it in the notation.
Within the random-phase approximation (RPA) we have the Dyson equation
(see, e.g., Ref.~\cite{MAH90}):
\begin{equation}
\Pi^{\mu\nu,\mu'\nu'}_{n,n'}=
\Pi^{0\mu\nu}_n\left(
\delta_{n,n'}\delta_{\mu,\nu'}\delta_{\nu,\mu'}
+\sum_{h,\mu'',\nu''} W_h^{\nu\mu''\nu''\mu}\Pi^{\mu''\nu'',\mu'\nu'}_{n+h,n'}
\right)
\label{Eqdysonpol}
\end{equation}
where the vacuum polarizability $\Pi^{0}$ is given by \cite{MAH90}
\begin{equation}
\Pi^{0\mu\nu}_n(\ul{p},\omega=0)
=\frac{2}{A}
\sum_{\ul{k}}\frac{n_F(E_{\ul{k}+\ul{p}}+E^{\nu}-E^F_n)-n_F(E_{\ul{k}}+E^{\mu}-E^F_n)}
{(E_{\ul{k}+\ul{p}}+E^{\nu})-(E_{\ul{k}}+E^{\mu})}\, .
\end{equation}
Now we find that $\Pi^{0bb}_n(\ul{p},\omega=0)=0$ assuming that the upper level
is not occupied.
$\Pi^{0ab}_n(\ul{p},\omega=0)$ is quite small as the gap appears
in the denominator. It remains the contribution 
$\Pi^{0aa}_n(\ul{p},\omega=0)$ which yields for the 2 dimensional 
electron gas \cite{STE67a}: 
\begin{equation}
\Pi^{0aa}_n(\ul{p},\omega=0)=-\frac{m}{\pi\hbar^2}\left(1-\Theta(p-2k_F)
\sqrt{1-4\frac{k_F^2}{p^2}}\right)
\end{equation}
which is independent of $p$ for $p<2k_F$. As all polarizations 
only have $a$-indices 
we omit these indices in the following. Eq.~(\ref{Eqdysonpol}) can be solved 
by a Fourier transformation (for an infinite superlattice and assuming
$\Pi^{0}_n=\Pi^{0}$ is independent of $n$). Defining 
$\tilde{\Pi}_q=\sum_{n'}\Pi_{n,n'}e^{iq(n'-n)}$, 
$\tilde{W}^{aaaa}_q=\sum_{h}W^{aaaa}_he^{iqh}$
we find 
\begin{equation}
\tilde{\Pi}_q=\frac{\Pi^{0}}{1-\Pi^{0}\tilde{W}_q^{aaaa}}\,.
\end{equation}
Now the screened electron-impurity interaction is given by
the bare interaction and a part combined with the polarizability given by
\begin{equation}
V_{i,n}^{aa\, {\rm  sc}}(\ul{p})=V_{i,n}^{aa}
+\sum_{h,n'}W^{aaaa}_{h}\Pi_{n+h,n'}V_{i,n'}^{aa}\, .
\end{equation}
Defining $\tilde{V}_{i,q}^{aa}=\sum_{n}V_{i,n}e^{-iq(n-n_i)}$
and using the translational invariance $\Pi_{n+h,n'}=\Pi_{2n+h-n',n}$ 
we find:
\begin{equation}
V_{i,n}^{aa\, {\rm sc}}(\ul{p})=
\frac{1}{2\pi}\int_{-\pi}^{\pi} dq\frac{\tilde{V}_{i,q}^{aa}}
{1-\Pi^{0}\tilde{W}^{aaaa}_q}e^{iq(n-n_i)} \label{EqVsca}
\end{equation}

Similarly we have:
\begin{eqnarray}
&&V_{i,n}^{bb\, {\rm sc}}(\ul{p})=V_{i,n}^{bb}
+\sum_{h,n'}W^{baab}_{h}\Pi_{n+h,n'}V_{i,n'}^{aa}\label{EqVscb} \\
&&\quad =\frac{1}{2\pi}\int_{-\pi}^{\pi}dq\left[
\frac{\tilde{V}_{i,q}^{bb}}
{1-\Pi^{0}\tilde{W}^{aaaa}_q}+
\frac{(\tilde{W}^{baab}_q\tilde{V}_{i,q}^{aa}-
\tilde{W}^{aaaa}_q\tilde{V}_{i,q}^{bb})\Pi^{0}}
{1-\Pi^{0}\tilde{W}^{aaaa}_q}\right]e^{iq(n-n_i)}\, .
\nonumber
\end{eqnarray}
The screened matrix elements from Eqs.~(\ref{EqVsca},\ref{EqVscb}) are
used in the Hamiltonian~(\ref{EqH0imp}) in the following.

\begin{figure}
\vspace*{2cm}
\includegraphics{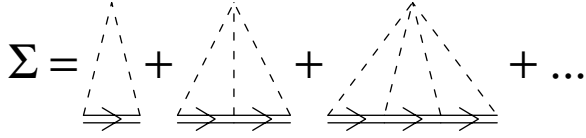}
\caption[a]{Diagrams contained in the self-consistent single-site approximation}
\label{Figssa}
\end{figure}

Now we calculate the self energy within the single-site approximation
which is shown diagrammatically in Fig.~\ref{Figssa}.
For the impurity $i$ we find the contribution to the self energy 
$\Sigma_{n}^{a\,{\rm ret}}(\ul{k},E)$
\begin{eqnarray}
\lefteqn{\Sigma^{a,i}_{n}(\ul{k},E)=\frac{1}{A^2}\sum_{\ul{k}_1}
V_{i,n}^{aa\, {\rm sc}}(\ul{k}-\ul{k}_1) G(\ul{k}_1,E)
V_{i,n}^{aa\, {\rm sc}}(\ul{k}_1-\ul{k})} \\
&&\quad +\frac{1}{A^3}\sum_{\ul{k}_1,\ul{k}_2}
V_{i,n}^{aa\, {\rm sc}}(\ul{k}-\ul{k}_1) G(\ul{k}_1,E)
V_{i,n}^{aa\, {\rm sc}}(\ul{k}_1-\ul{k}_2) G(\ul{k}_2,E)
V_{i,n}^{aa\, {\rm sc}}(\ul{k_2}-\ul{k})\nonumber \\
&&\quad +\frac{1}{A^4}\sum_{\ul{k}_1,\ul{k}_2,\ul{k}_3}
V_{i,n}^{aa\, {\rm sc}}(\ul{k}-\ul{k}_1) G(\ul{k}_1,E)
V_{i,n}^{aa\, {\rm sc}}(\ul{k}_1-\ul{k}_2) G(\ul{k}_2,E)
\nonumber \\ 
&&\quad \qquad \times V_{i,n}^{aa\, {\rm sc}}(\ul{k}_2-\ul{k}_3)G(\ul{k}_3,E)
V_{i,n}^{aa\, {\rm sc}}(\ul{k}_3-\ul{k})
+\dots\nonumber
\end{eqnarray}
where $G(\ul{k},E)=(E-E_k-\Sigma_{n}^{a\,{\rm ret}}(\ul{k},E))^{-1}$
is the full retarded Green function.
This sum can be transformed to the self-consistent equation 
(see, e.g., \cite{GOL88})
\begin{equation}
K^{a,i}(\ul{k}_1,\ul{k},E)=V_{i,n}^{aa\, {\rm sc}}(\ul{k}_1-\ul{k})
+\frac{1}{A}\sum_{\ul{k}_2}
V_{i,n}^{aa\, {\rm sc}}(\ul{k}_1-\ul{k}_2) G(\ul{k}_2,E)
K^{a,i}(\ul{k}_2,\ul{k},E)
\label{EqK}
\end{equation}
which can be solved numerically for a given self-energy function
$\Sigma_{n}^{a\,{\rm ret}}(\ul{k},E))$ entering $G(\ul{k}_2,E)$.
The contribution to the self energy is then given by
\begin{equation}
\Sigma^{a,i}_{n}(\ul{k},E)=
\frac{1}{A^2}\sum_{\ul{k}_1}
V_{i,n}^{aa\, {\rm sc}}(\ul{k}_1-\ul{k}) G(\ul{k}_1,E)
K^{a,i}(\ul{k}_1,\ul{k},E)\, .
\label{Eqsigmai}
\end{equation}
Summing up the contribution from all impurities $i$
and possibly different scattering processes we obtain the 
self-energy function 
\begin{equation}
\Sigma_{n}^{a\,{\rm ret}}(\ul{k},E)=\sum_i\Sigma^{a,i}_{n}(\ul{k},E)
+\Sigma^{a,{\rm other\, scattering}}_{n}(\ul{k},E)\, .
\label{Eqsigmatotal}
\end{equation}
Now a self consistent solution of 
Eqs.~(\ref{EqK},\ref{Eqsigmai},\ref{Eqsigmatotal}) can be achieved by 
iteration. 

The contribution to $\Sigma_{n}^{b\,{\rm ret}}(\ul{k},E))$
is calculated in the same way.

\subsection{Interface roughness}
Interface roughness is modelled like in Ref.~\cite{WACpb}
considering an interface located at $z=z_0$ exhibiting thickness 
fluctuations $\xi(\ul{r})$ of the order of
$\pm \eta$ (we use $\eta=2.8$ {\AA} which is one monolayer of GaAs). 
We assume the correlations
\begin{eqnarray}
\langle \xi(\ul{r}) \rangle_r&=&0\\
\langle \xi(\ul{r})\xi(\ul{r}') \rangle_r
&=&\alpha \eta^2\exp(-|\ul{r}-\ul{r}'|/\lambda)\label{Eqexpkorr}
\end{eqnarray} 
with a correlation length $\lambda=7$ nm and an average coverage
$\alpha=0.5$. Such an exponential correlation function 
$\langle \xi(\ul{r})\xi(\ul{r}') \rangle_r$ seems to be more
appropriate than the usual choice of a Gaussian (see, e.g., Ref.~\cite{DHA90})
as stated in Refs.~\cite{GOO85,WACpb}.
Like in Ref.~\cite{DHA90} (where the scattering of Bloch states
in a superlattice is regarded) the 
additional potential\footnote{In chapter 11 of this book the 
potential is chosen to be the variation of
the energy levels due to the well width fluctuation which is only defined
for scattering within a given level and a given well. 
The approach via Eq.~(\ref{Eqroughpot}) has the
advantage that interwell and interlevel transitions  can be handled as well.
For intrawell and intralevel processes the results are similar  
as d$E^{\nu}/$d$w\sim\Delta E_c |\Psi^{\nu}(z_0)|^2$ 
where $w$ is the well width and
$z_0$ is the position of the interface.}
due to the roughness is modelled 
by a $\delta$-function at the perfect interface
\begin{equation}
U(\ul{r},z)=\xi(\ul{r})\Delta E_c \delta(z-z_0)\, . \label{Eqroughpot}
\end{equation} 
This gives the following interface roughness contribution to
$\hat{H}^{\rm scatter}$
\begin{eqnarray}
\hat{H}^{\rm rough}&=&\frac{1}{A}
\sum_{\ul{k},\ul{p},h}\left[
U^{aa}_{h}(\ul{p})a^{\dag}_{n+h}(\ul{k}+\ul{p})a_{n}(\ul{k})
+U^{bb}_{h}(\ul{p})b^{\dag}_{n+h}(\ul{k}+\ul{p})b_{n}(\ul{k})\right. 
\nonumber \\
&&\left.
+U^{ba}_{h}(\ul{p})b^{\dag}_{n+h}(\ul{k}+\ul{p})a_{n}(\ul{k})
+U^{ab}_{h}(\ul{p})a^{\dag}_{n+h}(\ul{k}+\ul{p})b_{n}(\ul{k})
\right]\label{Eqhamrough}
\end{eqnarray}
with the matrix elements
\begin{equation}
U^{\nu\mu}_h(\ul{p})=
\int d^2r e^{-i\ul{p}\cdot\ul{r}}\Delta E_c
\left[\xi(\ul{r})\Psi^{\nu\, *}(z_0-hd)
\Psi^{\mu}(z_0)\right]\, .
\end{equation}
Using the correlation function (\ref{Eqexpkorr}) we obtain the square
of the matrix element
\begin{equation}
|U^{\nu\mu}_h(\ul{p})|^2=A \Delta E_c^2
|\Psi_{\nu}(z_0-hd)|^2|\Psi_{\mu}(z_0)|^2
\frac{2\pi\alpha \eta^2 \lambda^2}{\left(1+(p\lambda)^2\right)^{3/2}}
\label{Eqroughmat}
\end{equation}
which enters the expressions in the following.

The elements $U_0^{\nu\mu}$ result in scattering within the wells.
Their contribution to the self-energy is calculated within the
self-consistent Born approximation (which is just the first diagram
of the infinite sum in Fig.~\ref{Figssa})
\begin{equation}
\Sigma_{{\rm rough}}^{a\, {\rm ret}}(\ul{k},E)
=\frac{2}{A^2}\sum_{\ul{k}_1}
|U^{aa}_0(\ul{k}-\ul{k}_1)|^2 G(\ul{k}_1,E)\, ,
\label{Eqsigmarough}
\end{equation}
where the factor 2 takes into account the two interfaces per well.
These self-energies contribute to the total self energy in 
Eq.~(\ref{Eqsigmatotal}).
The calculation for the subband $b$ is performed in the same way.

The elements $U_1^{\nu\mu}$ contribute to the non-resonant 
current from  one well to the next
via Eq.~(\ref{EqJ}). Here the contributions from all 4 interfaces of both
wells involved are summed up.
For weakly coupled wells 
$U_2^{\nu\mu}, U_3^{\nu\mu}, \ldots$  are small and can be  
neglected.

\subsection{Optical phonons}
In polar materials like GaAs the polar interaction with optical phonons
provides an important scattering process.
As the energy $\hbar \omega_0=36$ meV is transferred, this process couples
electronic states with different energy in contrast to the two
scattering mechanism discussed before. This makes the full 
calculation of the self-energies much more complicated.
But fortunately only a restricted number of processes are allowed
at low temperatures where only the emission of phonons takes place.
Therefore to any electronic state $\ul{k}$ in level $\nu$ with energy
$E_k+E^{\nu}$ affected there
must be another state at an energy $E_k+E^{\nu}-\hbar \omega_0$.
For the lowest level $\nu=a$ this means that the condition 
$E_k>\hbar \omega_0$ must be satisfied
in order for phonon scattering to be possible.
But for small Fermi levels (5.37 meV for the sample considered)
these states are far away from any resonant transition
so that the neglection of the phonon contribution to the self-energy
hardly affects the currents.
The situation is different for the second level. Here the states 
with $E_k\approx 0$ are in resonance with the occupied states in the 
ground level if the electric field takes the value $eFd\approx E^b-E^a$.
Therefore the actual broadening of these states is crucial for
the $a\to b$ resonance.  This process is taken into account
by calculating the scattering time $\tau_{\rm ph}$ for this
process following Ref.~\cite{FER89} yielding 
$\tau_{\rm ph}=0.854$ psec for the structure considered here.
For the self-energy contribution the constant value
\begin{equation}
\Sigma_{{\rm phonon}}^{b\, {\rm ret}}(\ul{k},E)
=-i\frac{\hbar}{2\tau_{\rm ph}}
\end{equation}
is used which contributes to the total self energy in 
Eq.~(\ref{Eqsigmatotal}) for the level $b$.

\section{Conclusions}
In this chapter the electronic transport in weakly coupled multiple
quantum wells has been considered.
Within the model of sequential tunnelling the 
currents have been  calculated without any fitting parameters
taking into account the scattering at impurities and interface roughness.
The currents are in good quantitative agreement with experimental data
stating the physical relevance of the model.

In the NDC region a homogeneous field distribution is unstable.
For the actual doping of the sample considered both
stable field-domains and self-sustained current oscillations
are found theoretically in good agreement with the experimental observation.
The sequence of domain branches is almost independent
of the contacts as the influence of the boundaries
is shielded by the domain regions where the electric field 
is in the regime of positive differential conductivity.
Therefore the branches contain information about
the transport  in multiple quantum wells itself 
which is not spoiled by contacts which are often only poorly defined.
Furthermore the domain branches react quite sensitively
to local deviations from periodicity which allows for a check
of the actual sample quality.

I want to thank Luis Bonilla, Holger Grahn, Ben Hu, Anatoli Ignatov, 
Antti-Pekka Jauho, Kristinn Johnsen, J{\"o}rg Kastrup, 
Miguel Moscoso, Michael Patra, 
Frank Prengel, Eckehard Sch{\"o}ll, Georg Schwarz, and Stefan Zeuner 
for fruitful collaboration  and helpful discussions.
Financial support from the Deutsche Forschungsgemeinschaft
is gratefully acknowledged.


\end{document}